\documentclass[twocolumn,aps,prl,reprint]{revtex4-2}

\usepackage[english]{babel}
\usepackage{titlesec}
\usepackage{ragged2e}
\usepackage[normalem]{ulem}
\usepackage{afterpage}
\usepackage{soul,xcolor}

\newcommand{\vq}{\mathbf{q}}
\renewcommand{\vr}{\mathbf{r}}

\usepackage{amsfonts, amsmath, amssymb, amsthm}

\usepackage{graphicx}
\usepackage{float}
\usepackage{caption}
\usepackage{subcaption}
\usepackage{enumitem}

\usepackage{makecell}
\usepackage{multirow}
\usepackage{booktabs}

\usepackage{changes}
\usepackage{comment}
\soulregister\ref{7}
\soulregister\textit{1}

\usepackage[colorlinks=true]{hyperref}

\newcommand{\br}[1]{\left( #1 \right)}
\newcommand{\sbr}[1]{\left[ #1 \right]}
\newcommand{\abr}[1]{\left\langle #1 \right\rangle}
\newcommand{\abs}[1]{\left\vert #1 \right\vert}

\newcommand{\iu}{{\hat{i}\mkern1mu}}

\begin{document}

\title{Effective hyperuniformity in time-integrated stochastic Turing patterns}

\author{Anirban Mukherjee$^{1}$}
\email{anirban9973@gmail.com}
\author{Hong-Yan Shih$^{1,2}$}
\email{hongyan@as.edu.tw}
\affiliation{%
 $^1$Institute of Physics, Academia Sinica, Taipei 115201, Taiwan\\
 $^2$Physics Division, National Center for Theoretical Sciences, Taipei 106319, Taiwan
}

\begin{abstract}
Demographic noise generates stochastic Turing patterns even when
reaction-diffusion systems are deterministically stable. We show
analytically and verify numerically in the Levin-Segel model that
temporal integration of configurations reveals emergent large-scale
organization. The intensive number variance in a window of size
$R \gg 1$ approaches a finite reaction-kinetic floor as $1/R$, over a
spatial range growing by orders of magnitude near the Turing
instability. This yields an effectively hyperuniform,
fine-tuning-free regime previously unidentified in non-conserved
multispecies stochastic systems.
\end{abstract}

\maketitle

\textit{Introduction.}---
Living systems often display striking spatial patterns---from stripes and
spots in animal pigmentation to vegetation bands in drylands---raising a
fundamental question: how does robust macroscopic organization emerge at
scales far larger than those of individual
constituents~\cite{Murray2003, Cross1993Jul, Camazine2003Sep,
Rietkerk2021Oct}? Turing's landmark theory showed that such patterns can
emerge spontaneously from a homogeneous state through the coupling of a
slowly diffusing activator to a faster-diffusing inhibitor, a central
paradigm of reaction--diffusion pattern formation~\cite{Turing1952Aug,
Kondo2010Sep, Rietkerk2008Mar}. Central to this picture, particularly in
developmental biology~\cite{Weber2022Mar}, are
\textit{morphogens}~\cite{Smith1996May, Gurdon2001Oct, Wolpert1996Sep},
signaling molecules whose local concentration acts as a positional cue for
how cells differentiate and organize. In most such systems noise is
intrinsic and unavoidable~\cite{Tsimring2014Jan, lande2003}: stochastic
reaction--diffusion models capture this discreteness through individual
reaction and diffusion events, generating finite-wavelength patterns even
below the deterministic Turing threshold by selectively amplifying weakly
damped modes at an intermediate wave number~\cite{McKane2004Oct,
Lugo2008Nov, McKane2005Jun, Butler2009Sep, Butler2011Jul, Biancalani2010Apr,
Asslani2012Oct}, a mechanism invoked for patterning in synthetic bacterial
populations~\cite{Karig2018Jun}, cyanobacterial
filaments~\cite{DiPatti2018May}, and Arabidopsis
trichomes~\cite{DiPatti2023Oct} (for a review, see
Ref.~\cite{DiPatti2026Jun}).

The prevailing treatment assumes that biological entities read the local
morphogen concentration at a given instant~\cite{Turing1952Aug,
Wolpert1969Oct, Pages2000Jan, Ashe2006Feb}, as do activator-inhibitor models
of ecological pattern formation~\cite{Murray2003, Rietkerk2008Mar,
Borgogno2009Mar}. Yet cells and organisms often do not act on instantaneous
snapshots alone; in several patterning systems they accumulate signals and
make fate decisions reflecting a history of morphogen~\cite{Pages2000Jan,
Harfe2004Aug, Dessaud2007Nov, Dessaud2008Aug, Sagner2017Jul} and
stressor~\cite{Ogle2015Mar, Johnstone2016Sep} activity rather than---or in
addition to---the instantaneous concentration. This raises a question: can
large-scale spatial organization emerge from accumulated fluctuations,
beyond the instantaneous patterns at intermediate Turing scales?

A particularly stringent form of large-scale spatial organization is
hyperuniformity, in which long-wavelength density fluctuations are
anomalously suppressed: the number variance grows more slowly than the
observation volume, or equivalently the structure factor vanishes in the
long-wavelength limit, $S(\vq) \rightarrow 0$ as $|\vq| \to
0$~\cite{Torquato2003Oct, Torquato2018Jun}. Beyond crystals and
quasicrystals, hyperuniformity also occurs in disordered systems that
appear irregular locally but highly uniform at large scales, and has been
identified in biological and ecological systems where such uniformity is
often linked to function, including the avian retinal photoreceptor
mosaic~\cite{Jiao2014Feb}, looped leaf-vein networks~\cite{Liu2024Jul},
dryland vegetation~\cite{Ge2023Oct}, swimming algae~\cite{Huang2021May},
immune receptors~\cite{Mayer2015May}, and epithelial tissue
layers~\cite{Tang2024, Siegert2026May}. Deterministic pattern-forming
equations, including the Swift--Hohenberg and Cahn--Hilliard
equations~\cite{Ma2017Jun, Zheng2024Sep} and generic Gray--Scott
models~\cite{Ballestero2025Dec}, can also generate hyperuniform states. In
equilibrium systems, hyperuniformity arises naturally in crystals,
quasicrystals, and certain disordered ground states with long-range
interactions or constrained correlations, whereas in stochastic
nonequilibrium systems known routes to hyperuniformity are more restrictive
and often rely on conservation laws: particle-number conservation can
suppress long-wavelength density fluctuations~\cite{Hexner2015Mar,
Mukherjee2023Feb}, while conservation of the center of
mass~\cite{Hexner2017Jan, Mukherjee2024Aug} or higher moments of the mass
distribution~\cite{Maire2025Dec} enforces stronger forms of
hyperuniformity. Although hyperuniform density fluctuations were recently
reported in a nonconserved stochastic population model through
self-organized criticality~\cite{agranov2025}, whether noisy
reaction--diffusion systems can generically produce hyperuniform-like order
without conservation laws or imposed constraints remains unresolved.

Here we show that time integration can reveal effective hyperuniform
organization in stochastic reaction--diffusion systems exhibiting
finite-wavelength Turing patterns. Time integration acts as a dynamical
filter for persistent fluctuations, so that the zero-frequency dynamic
structure factor at small wavenumber and the long-time intensive subsystem
fluctuations at large scales are controlled by a noise floor pinned by
reaction kinetics. In the stochastic Turing regime, this reaction-induced
floor is subdominant over an intermediate range of length scales above the
Turing scale, taking over only beyond a crossover length that diverges as
the Turing instability is approached. The resulting suppression of
long-wavelength fluctuations reflects an incomplete but increasingly
extended spatiotemporal compensation: local clustering promotes growth but
also enhances inhibition, generating anticorrelated fluctuations whose spatial
sum leaves only a finite reaction-controlled residual. Near the instability,
the weak damping of the Turing modes sharpens the pattern and extends the
oscillatory correlations, making this compensating response long-ranged and
long-lived enough to let the hyperuniform-like regime dominate the floor
across a broad range of scales. Thus, although particle number is not
conserved, reactions impose an emergent constraint on accumulated-density
fluctuations, providing a nonconserved route to effective hyperuniformity.
Our stochastic simulations confirm the theory and show that temporal
accumulation sharpens spatial structures nearly invisible in instantaneous
snapshots, suggesting effective hyperuniformity as a diagnostic for hidden
order in stochastic pattern-forming systems.
%%%%%%%%%%%%%%%%%%%%%%%%%%%%%%%%%%%%%%%%%%%%%%%%%%%%%%%%%%%%%%%%%%%%%%%%%%%%%%%%

%%%%%%%%%%%%%%%%%%%%%%%%%%%%%%%%%%%%%%%%%%%%%%%%%%%%%%%%%%%%%%%%%%%%%%%%%%%%%%%%
\textit{Time-integrated configurations.---}We consider a generic
two-species stochastic reaction-diffusion process on a discrete
$d$-dimensional hypercubic lattice with periodic boundary conditions.
Each site $\vr$ has internal volume $V$ and hosts integer populations
$N^{(X)}_{\vr}(t)$, with $X = P$ (activator) or $Q$ (inhibitor).
Using the van Kampen system-size expansion~\cite{VanKampen2007,
Lugo2008Nov}, we decompose the local populations as
$N_{\vr}^{(X)} \simeq V \rho_{\vr}^{(X)} + \sqrt{V} \xi_{\vr}^{(X)}$.
The deterministic part $\rho_{\vr}^{(X)}$ obeys the mean-field
reaction-diffusion equation
\begin{equation}
    \partial_t \rho_{\vr}^{(X)} = f_X\!\left(\boldsymbol{\rho}_{\vr}\right)
    + \sum_{Y} \mathcal{D}_{XY} \nabla^2 \rho_{\vr}^{(Y)},
    \label{eq:deterministic_RD}
\end{equation}
where $f_X$ is the local reaction kinetics and $\mathcal{D}$ is the
diffusion matrix coupling species $X$ and $Y$. The fluctuating part
$\xi_{\vr}^{(X)}$ obeys the linear Langevin equation
\begin{equation}
    \partial_t \xi_{\vr}^{(X)} = \sum_{Y} \mathcal{R}_{XY}\, \xi_{\vr}^{(Y)}
    + \sum_{Y} \mathcal{D}_{XY} \nabla^2 \xi_{\vr}^{(Y)} + \eta_{\vr}^{(X)}(t),
    \label{eq:langevin}
\end{equation}
where $\mathcal{R}$ is the reaction matrix, obtained by linearizing
$f_X$ about the homogeneous steady state, and is the same matrix that
enters the standard deterministic Turing stability analysis. The
noise $\eta_{\vr}^{(X)}(t)$ is Gaussian and white,
$\langle \eta_{\vr}^{(X)}(t)\, \eta_{\vr'}^{(Y)}(t') \rangle =
B^{(XY)}(\boldsymbol{\rho}) \,\delta_{\vr\vr'}\, \delta(t-t')$, and
encodes the demographic noise from the underlying
reactions~\cite{SuppMat}. As $V \to \infty$ at fixed time, $\xi_\vr^{(X)}$
converges to the Gaussian process governed by
Eq.~\eqref{eq:langevin}~\cite{Kurtz1970Apr,Kurtz1971Jun}, together
with Eq.~\eqref{eq:deterministic_RD} constituting the linear noise
approximation (LNA).

%% Description about the deterministic turing pattern condition
Deterministic Turing pattern formation requires the underlying
reaction-diffusion system to be unstable~\cite{Turing1952Aug, Cross2009Jul}
for at least one wave vector $\vq^*$, via the condition
\begin{equation}
    \label{eq:deterministic_Turing_condition}
    \det \mathcal{A}_{\vq^*} = 0,
\end{equation}
where the stability matrix is given by
$\mathcal{A}_\vq = \br{\mathcal{R} - \mathcal{D} \abs{\vq}^2}$;
this requires the inhibitor-to-activator diffusivity ratio to exceed
a threshold set by $\mathcal{R}$.
Crucially, the Turing mechanism also requires the
well-mixed system itself to be stable,
$\det\mathcal{A}_{\vq=0} = \det\mathcal{R} > 0$, so any Turing
instability necessarily occurs at a finite wave vector
$\abs{\vq^*}>0$, leaving the $\vq\to0$ limit governed entirely by
stable dynamics at the deterministic level.
The system remains stable and does not form any Turing pattern if the
stability matrix has a positive determinant, $\det \mathcal{A}_\vq > 0$,
and a negative trace, $\operatorname{tr}\,\mathcal{A}_\vq < 0$,
for all wave vectors $\vq$.

%%%%%%%%%%%%%%%%%%%%%%%%%%%%%%%%
%Stochatic turing pattern setup
%%%%%%%%%%%%%%%%%%%%%%%%%%%%%%%%
Stochastic Turing patterns can develop~\cite{Butler2009Sep, Butler2011Jul}
in an otherwise stable deterministic reaction--diffusion system when
demographic fluctuations, arising from the intrinsic stochasticity of
the dynamics and amplified by the slowly relaxing eigenmodes of the
stability matrix~\cite{McKane2005Jun}, produce a peak for a certain
wave vector $\vq^*$ at \textit{zero frequency} in the dynamic structure factor of $\xi_{\vr}^{(X)}(t)$.
Unlike deterministic Turing patterns, stochastic Turing patterns are not
static in steady state: the \textit{statistics} (mean densities, dynamic
power spectra, and equal-time structure factors) are stationary, but
individual configurations $N_{\vr}^{(X)}(t)$ fluctuate indefinitely,
mixing fast reaction noise with slow diffusion-mediated spatial
organization, even producing giant fluctuations~\cite{Biancalani2017Jan}.

Our main objects of interest are the time-integrated demographic fluctuations,
\begin{equation}
    \Xi^{(X)}_\vr(T) = \int_0^T dt\, \xi^{(X)}_\vr(t),
\end{equation}
and their structure factor in steady state,
\begin{equation}
    S^{(X)}_{\vq}(T) = \frac{\abr{\tilde{\Xi}_{\vq}^{(X)}(T)\,
    \tilde{\Xi}_{-\vq}^{(X)}(T)}}{L^d T},
\end{equation}
where $\tilde{\Xi}_{\vq}^{(X)}(T)$ is the spatial Fourier transform
of $\Xi^{(X)}_\vr(T)$. In the limit $T \to \infty$, the structure
factor equals the zero-frequency component of the dynamic power
spectrum of species $X$,
\begin{equation}
\label{eq:structure_factor_time_integrated_configs}
    \overline{S}^{(X)}_{\vq} \equiv \lim_{T \rightarrow \infty}
    S^{(X)}_{\vq}(T) = \abr{\tilde{\xi}_{\vq}^{(X)}(0)
    \tilde{\xi}_{-\vq}^{(X)}(0)},
\end{equation}
where $\tilde{\xi}^{(X)}_\vq(\omega)$ denotes the spatio-temporal
Fourier transform of $\xi^{(X)}_\vr(t)$.
The convergence timescale of $S^{(X)}_{\vq}(T)$ to
$\overline{S}^{(X)}_{\vq}$ is model-dependent and physically
meaningful for natural systems, as it quantifies the integration
time required to resolve the large-scale spatial organization of
time-integrated configurations.

%% Large scale strcuture of stochastic Turing patterns
\textit{Large-scale structure.---}The large-scale structure is encoded in the small-wavevector behavior as
\begin{equation}
  \label{eq:final_zero_omega_small_q_spectrum}
    \overline{S}^{(X)}_{\vq} \simeq \mathcal{F}^{(X)} +
    \mathcal{G}^{(X)} \abs{\vq}^{2}
    + \mathcal{O}(\abs{\vq}^{4}).
\end{equation}
$\mathcal{F}^{(X)}=\overline{S}^{(X)}_{\mathbf 0} > 0$ measures fluctuations of the total space-time-integrated population of species $X$ and is determined solely by the reaction kinetics, independent of diffusion~\cite{SuppMat}.
Because the $\vq=0$ mode remains linearly stable at the
deterministic level (see above; $\det\mathcal{A}_{\vq=0} =
\det\mathcal{R} > 0$), $\mathcal{F}^{(X)}$ has no deterministic
counterpart. It vanishes identically when the noise term
$\eta^{(X)}_{\vr}$ is switched off, rather than persisting as an
unavoidable floor of the structure factor.
Relatedly, in contrast to conserved stochastic systems, $\mathcal{F}^{(X)}$ is not
constrained to vanish by conservation laws or
fluctuation-dissipation relations~\cite{Mukherjee2023Feb, Mukherjee2024Aug}. Furthermore,
\begin{equation}
\label{eq:pattern_formation_condition}
\mathcal{G}^{(X)} > 0
\end{equation}
is the condition~\cite{Butler2011Jul} for stochastic Turing pattern
formation --- achievable even when the inhibitor-to-activator
diffusivity ratio lies below the deterministic threshold --- causing
$\overline{S}^{(X)}_{\vq}$ to attain a global maximum at a finite
wavevector $\vq^*$ and exhibit the small-wavevector scaling
$\overline{S}^{(X)}_{\vq}-\mathcal{F}^{(X)} \sim \abs{\vq}^{2}$
for $\abs{\vq}\ll \abs{\vq^*}$. It is also known that stochastic Turing patterns exhibit a fat-tail spectrum governed by $\overline{S}^{(X)}_{\vq}\sim \abs{\vq}^{-2}$ for $\abs{\vq}\gg \abs{\vq^*}$~\cite{Butler2011Jul}.
Beyond the leading quadratic term in
Eq.~\eqref{eq:final_zero_omega_small_q_spectrum}, higher-order
corrections proportional to $|\vq|^4, |\vq|^6, \ldots$ are also present.
Although their net effect is positive at small $|\vq|$, a negative
higher-order contribution emerges near $|\vq^*|$, causing the spectrum
to decrease beyond the Turing peak.

\textit{Subsystem fluctuations.}---
We quantify large-scale spatial organization of time-integrated
configurations through the population fluctuation within a spherical
window of radius $R$ centered at $\vr_0$,
\begin{equation}
    \Xi^{(X)}(R;T) = \sum_\vr \Xi^{(X)}_\vr(T)\, w(\vr-\vr_0;R),
\end{equation}
where $w(\vr-\vr_0;R) = 1$ if $\vr \in \Omega(\vr_0;R) \equiv \{\vr : |\vr-\vr_0|\leq R\}$
and $0$ otherwise, and $|\Omega(R)| = \sum_\vr w(\vr-\vr_0;R)$ denotes
the number of lattice sites inside the window.
In homogeneous steady state, the intensive population fluctuation
within that window at an arbitrary integration time $T$ is independent
of $\vr_0$ and can be written as
\begin{equation}
\label{eq:def_intensive_variance}
    {\rm var}^{(X)}(R;T) = \frac{1}{v_1(R)\,T}
    \abr{\sbr{\Xi^{(X)}(R;T)}^2},
\end{equation}
where $v_1(R) = \pi^{d/2} R^d / \Gamma(d/2+1)$ is the continuum
hyperspherical window volume, equal to $|\Omega(R)|$ for $R\gg 1$.
In the large-$T$ limit, $\overline{\rm var}^{(X)}(R) = \lim_{T \rightarrow \infty}
    {\rm var}^{(X)}(R;T)$,
can be expressed in terms of the structure factor as~\cite{SuppMat}
\begin{align}
\overline{\rm var}^{(X)}(R) = \frac{1}{L^d}
\sum_{\vq}\, \overline{S}^{(X)}_{\vq}\; \tilde{\alpha}_\vq(R).
\label{eq:var_subsystem}
\end{align}

\textit{The sum rule.}---In the large-$R$ limit, the window function
can be written as~\cite{SuppMat}
\begin{equation}
    \tilde{\alpha}_\vq(R) \xrightarrow{R\gg 1}
    \frac{2^d\pi^{d/2}\Gamma(d/2+1)}{\abs{\vq}^d}\,J_{d/2}^2(\abs{\vq}R) \geq 0,
    \label{eq:alpha_window}
\end{equation}
where $J_{d/2}$ is the Bessel function of the first kind of order $d/2$.
In the thermodynamic limit, $\lim_{R\to\infty}\tilde{\alpha}_\vq(R) =
(2\pi)^d\delta^d(\vq)$, so Eq.~\eqref{eq:var_subsystem} gives
\begin{equation}
    \lim_{R\to\infty}\lim_{L\to\infty}\overline{\rm var}^{(X)}(R)
    = \overline{S}^{(X)}_{\mathbf{0}} = \mathcal{F}^{(X)},
    \label{eq:total_fluctuations}
\end{equation}
and the corresponding sum rule on the correlation function of
time-integrated configurations,
\begin{equation}
    \sum_{\vr}\,\overline{C}_{\vr}^{(XX)} = \mathcal{F}^{(X)},
    \label{eq:sum_rule}
\end{equation}
where $\overline{C}_{\vr}^{(XX)} = \lim_{T\to\infty}T^{-1}
\abr{\Xi_{\mathbf{0}}^{(X)}(T)\,\Xi_{\vr}^{(X)}(T)}$.
The large-subsystem fluctuation is thus determined entirely by local
reaction kinetics, independent of diffusion and spatial organization
— a nonequilibrium counterpart of the grand-canonical
fluctuation-compressibility relation, in which the role of the
chemical reservoir is played by the reactions.

\textit{Effective hyperuniformity.}---The sum rule requires
$\overline{\rm var}^{(X)}(R)\to\mathcal{F}^{(X)}$ from above as
$R\to\infty$. Since the stochastic Turing peak forces
$\smash[t]{\overline{\rm var}}^{(X)}(R^*) > \mathcal{F}^{(X)}$
at $R^* \sim 1/|\vq^*|$, the variance must decay for $R > R^*$
to satisfy Eq.~\eqref{eq:total_fluctuations} — this decay is an
exact and inevitable consequence of the sum rule. The rate of decay
is controlled by the small-$|\vq|$ structure factor,
$\overline{S}_{\vq}^{(X)} - \mathcal{F}^{(X)} =
\mathcal{G}^{(X)}\abs{\vq}^2 + \mathcal{O}(\abs{\vq}^4)$
(cf.\ Eq.~\eqref{eq:final_zero_omega_small_q_spectrum}), signaling a
subleading class-I hyperuniform
correction~\cite{Torquato2003Oct, Torquato2018Jun}
\begin{equation}
    \overline{\rm var}^{(X)}(R) \approx \mathcal{F}^{(X)} +
    \frac{\mathcal{G}^{(X)}_{\rm eff}}{R},
\end{equation}
where $\mathcal{G}^{(X)}_{\rm eff}$ includes the higher-order
contributions from Eq.~\eqref{eq:final_zero_omega_small_q_spectrum}.
For $R^* \ll R \ll R_c^{(X)}$, where
$R_c^{(X)} = \mathcal{G}^{(X)}_{\rm eff}/\mathcal{F}^{(X)}$,
the subsystem variance effectively decays as
\begin{equation}
    \overline{\rm var}^{(X)}(R) \approx
    \frac{\mathcal{G}^{(X)}_{\rm eff}}{R}.
\end{equation}
As the system approaches the deterministic Turing instability
--- equivalently, as the inhibitor-to-activator diffusivity ratio
approaches its deterministic threshold from below ---
$\det\mathcal{A}_{\vq^*} \rightarrow 0$,
the eigenmode at $\vq^*$ becomes the slowest decaying and
demographic noise is maximally amplified, causing the spectral
peak to diverge as
\begin{equation}
  \label{eq:qualitative_strength_of_peak_height}
  \overline{S}_{\vq^*}^{(X)} \sim
  \frac{1}{\sbr{\det \mathcal{A}_{\vq^*}}^2}.
\end{equation}
Via Eq.~\eqref{eq:var_subsystem}, this divergence drives up
$\mathcal{G}_{\rm eff}^{(X)}$, and consequently $R_c^{(X)}$ increases, since $\mathcal{F}^{(X)}$ remains fixed by the reactions. In this regime, $R^*$ and $R_c^{(X)}$
become well separated, extending the window over which
$\overline{\rm var}^{(X)}(R)$ exhibits a $1/R$ decay across
several orders of magnitude in $R$. The system is therefore
effectively hyperuniform within this window.

%%%%%%%%%%%%%%%%%%%%%%%%%%%%%%%%%%%%%%%%%%%%%%%%%%%%%%%%%%%%%%%%%%%%%%%%%%%%%%%%%%%%%%%%%%%%%%%%%%%%%%%%%%%%%%%%%%%%%%%%%%%%%%%%%%%%%%%%%%%%%%%%%%%%%

\begin{figure*}[!htb]
    \includegraphics[width=\linewidth]{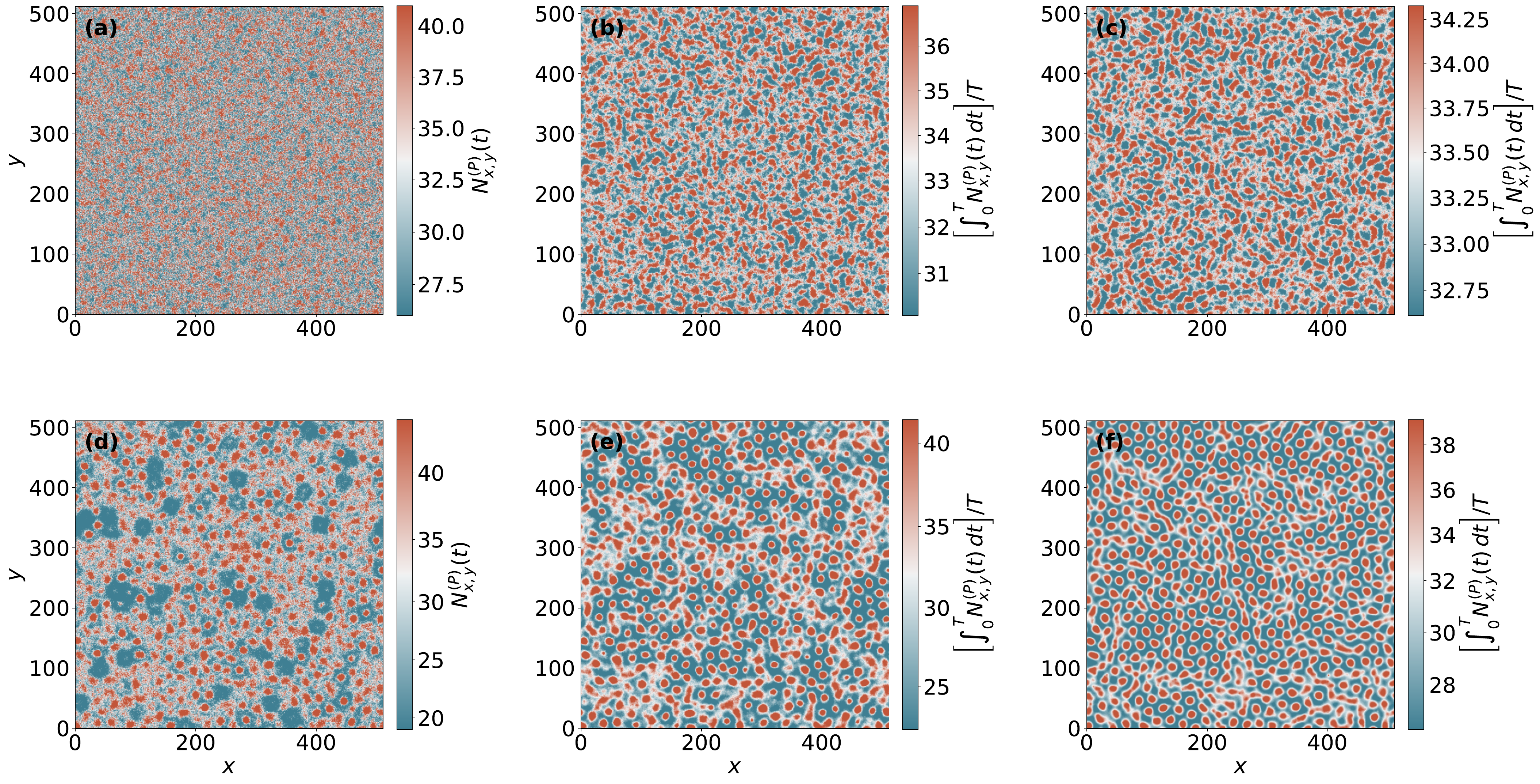}
    \caption{Spatial patterns of prey in the 2D Levin-Segel model on a
    $512 \times 512$ grid ($b = e = d = 0.5$, $p = 1$,
    $\mu = 1$, $V = 100$). (a--c) $\nu/\mu = 20$; (d--f) $\nu/\mu = 26$.
    Left panels (a,d): instantaneous steady-state configurations;
    middle and right panels (b,e) and (c,f): time-averaged over
    $T = 32$ and $T = 1024$ respectively, revealing how the spatially
    organized population structure gradually emerges as the integration
    time increases, absent in any single snapshot.
    The 2D simulation is shown for visual purposes only; all
    quantitative analyses are performed on the 1D model.}
    \label{fig:population_configuration_in_patterned_region_no_capacity}
\end{figure*}

\begin{figure}
    \centering
    \includegraphics[width=\linewidth]{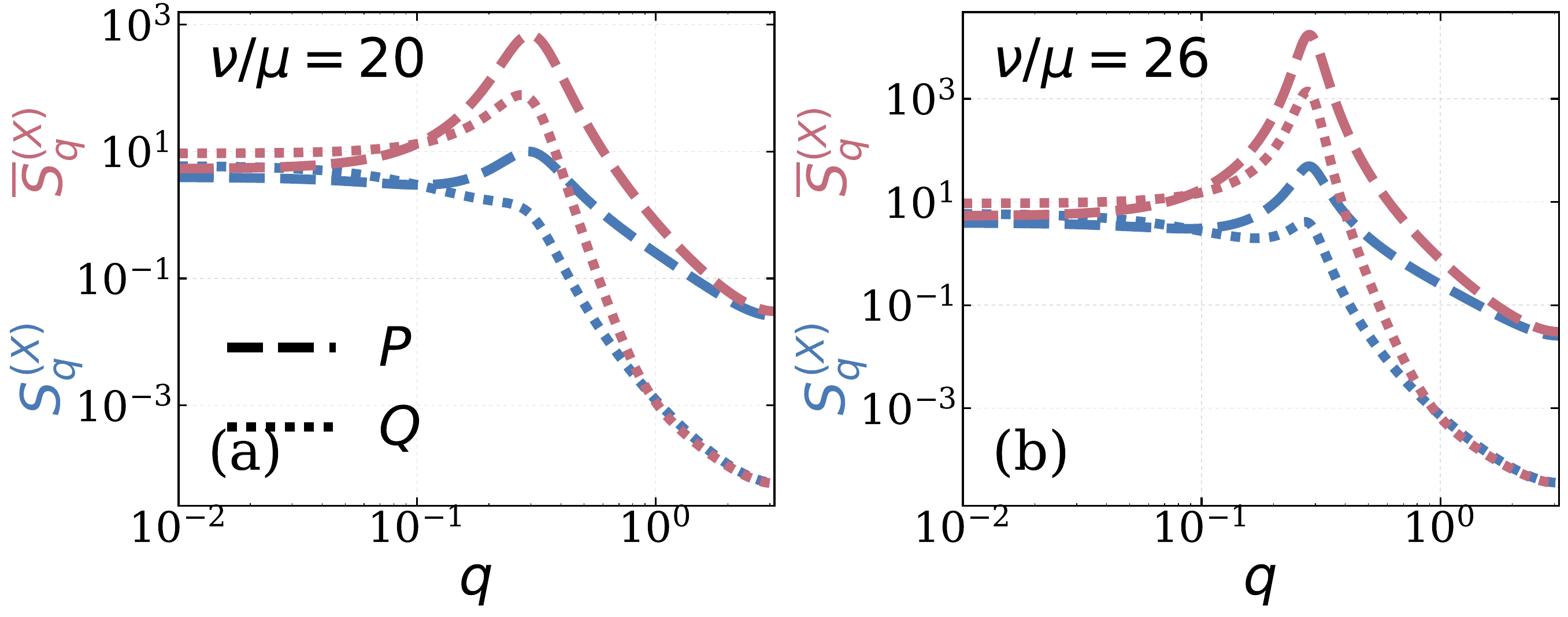}
    \caption{Instantaneous $S_q^{(X)}$ and time-integrated
    $\overline{S}_q^{(X)}$ structure factors for prey and
    predator at $\nu/\mu = 20$ (a) and $26$ (b). A peak signals the
    appearance of a spatial pattern: at $\nu/\mu = 20$, the predator
    shows no discernible peak instantaneously, and it only emerges
    after time integration, while the prey already shows a peak in
    both, but with a stronger one after time integration.}
    \label{fig:Sq_inst_vs_int}
\end{figure}

\textit{Stochastic Levin-Segel model.}---
We verify the existence of effective hyperuniformity in the stochastic
Levin-Segel predator-prey model~\cite{Levin1976Feb, Butler2009Sep}
on a ring.
The system consists of prey ($P$) and predator ($Q$) with onsite
reactions: prey birth $P \to 2P$ (rate $b$), Allee effect $2P \to 3P$ (rate $e/V$), predation $P + Q \to 2Q$
(rate $p/V$), and predator competition $2Q \to Q$ (rate $d/V$),
with diffusion rates $z\mu$ and $z\nu$ respectively for prey and predator, where $z$ is the
coordination number.
For reaction parameters $b = e = d = 0.5$ and $p = 1$, stochastic
Turing patterns form when $\nu/\mu > 2.48$, whereas the
deterministic instability requires $\nu/\mu > 27.8$~\cite{Butler2011Jul}.
In simulations, we use the next-reaction method~\cite{Gibson2000Mar} of
the Gillespie algorithm~\cite{Gillespie2007May, Gillespie2013May} and $V=1000$.
Fig.~\ref{fig:population_configuration_in_patterned_region_no_capacity}
shows that, in steady state, temporal integration up to $T=1024$
reveals markedly more regular spatial structure than instantaneous
snapshots in the prey (activator) population (see~\cite{SuppMat} for
predator (inhibitor) configurations). Although computed for one
dimension, the structure factor in Fig.~\ref{fig:Sq_inst_vs_int} makes
this quantitative, as a peak in the structure factor suggests the
appearance of stochastic Turing patterns. At $\nu/\mu = 20$
(Fig.~\ref{fig:Sq_inst_vs_int}(a)), the instantaneous predator
structure factor $S_q^{(Q)}$ shows no discernible peak; only the
time-integrated $\overline{S}_q^{(Q)}$ does, confirming that
temporal integration reveals spatial organization invisible in any
snapshot.

\begin{figure}[!htb]
    \centering
    \includegraphics[width=\linewidth]{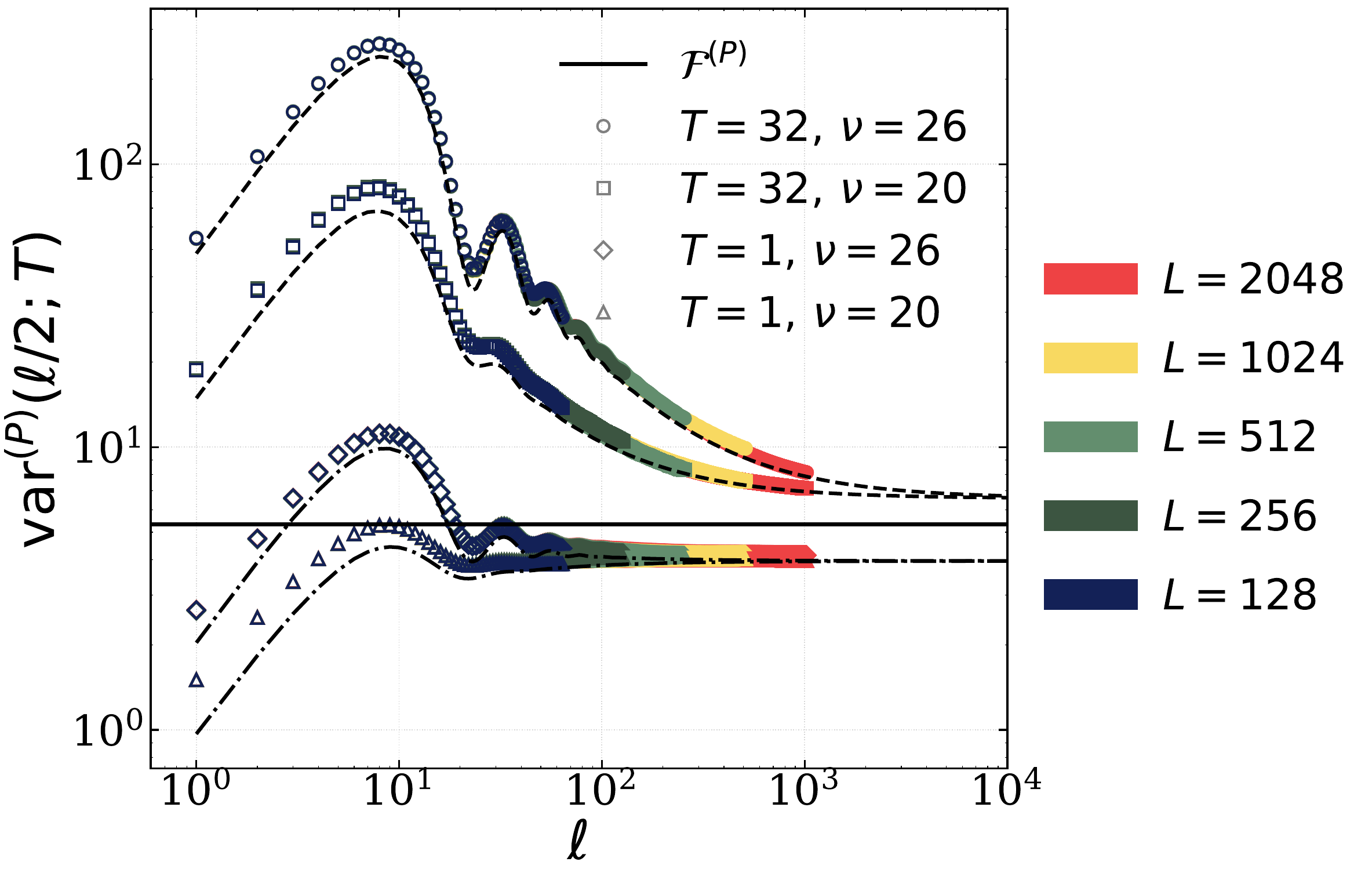}
    \caption{Prey subsystem variance ${\rm var}^{(P)}(\ell/2;\,T)$ versus
    subsystem size $\ell$, at $\nu/\mu = 20$ (away from the Turing
    instability) and $\nu/\mu = 26$ (near it), for integration times
    $T = 1$ and $T = 32$. Symbols: simulation data; dashed lines: LNA
    predictions. At $T=32$, fluctuations are enhanced near the Turing
    scale $\ell^*$ and suppressed beyond it --- the signature of
    large-scale spatial organization absent at $T=1$. The long-time
    saturation $\overline{{\rm var}}^{(P)}(\ell/2) \to \mathcal{F}^{(P)}$
    as $\ell \to \infty$ (solid black line) lies between the saturation
    values of $T=1$ and $T=32$, reflecting the nonmonotonic behavior
    of $\mathcal{F}^{(P)}(T)$ with $T$.}
    \label{fig:prey_T32_T1}
\end{figure}

A peak or a maximum in the structure factor at $q = q^*$
translates into a maximum of the scaled subsystem fluctuation,
${\rm var}^{(X)}(\ell/2;T)$, at the corresponding Turing length
scale $\ell^* \sim 1/\abs{\vq^*}$~\footnote{Here $R \equiv \ell/2$
denotes the hyperspherical sampling-window radius in general
dimension, reducing in our 1D simulations to half the subsystem
length $\ell$}. A higher peak
amplitude in the time-integrated structure factor must therefore
also produce a higher maximum in ${\rm var}^{(X)}(\ell/2;T)$. We
verify this for two diffusivity ratios, $\nu/\mu = 20$ (away from
the instability) and $\nu/\mu = 26$ (near the instability). As shown
in Fig.~\ref{fig:prey_T32_T1}, for both ratios the peak in
${\rm var}^{(P)}(\ell/2;T)$ at $\ell = \ell^*$ grows substantially
from $T=1$ to $T=32$, mirroring the growth of the corresponding
peak in the structure factor (Fig.~\ref{fig:Sq_inst_vs_int}). Beyond
this growth in peak height, time integration also reveals a
qualitatively new feature: at $T=1$, ${\rm var}^{(P)}(\ell/2;1)$
saturates quickly to its $\ell \gg 1$ value once $\ell > \ell^*$,
whereas at $T=32$ a regime of suppressed fluctuations emerges for
$\ell > \ell^*$ --- the signature of large-scale spatial
organization --- whose saturation value at $\ell \gg 1$ is
accessible only by systematically increasing $L$ from $128$ to
$2048$, since periodic boundary conditions restrict the maximum
subsystem size to $\ell = L/2$. Notably, the long-time ($T=4096$)
saturation value $\mathcal{F}^{(P)} = \overline{\rm var}^{(P)}(\ell/2)$
for $\ell \gg 1$ lies between the $T=1$ and $T=32$ curves,
demonstrating the nonmonotonic behavior of $\mathcal{F}^{(P)}(T)$
with $T$. See~\cite{SuppMat} for the corresponding variance of the
time-integrated predator population.

\begin{figure}
    \centering
    \includegraphics[width=\linewidth]{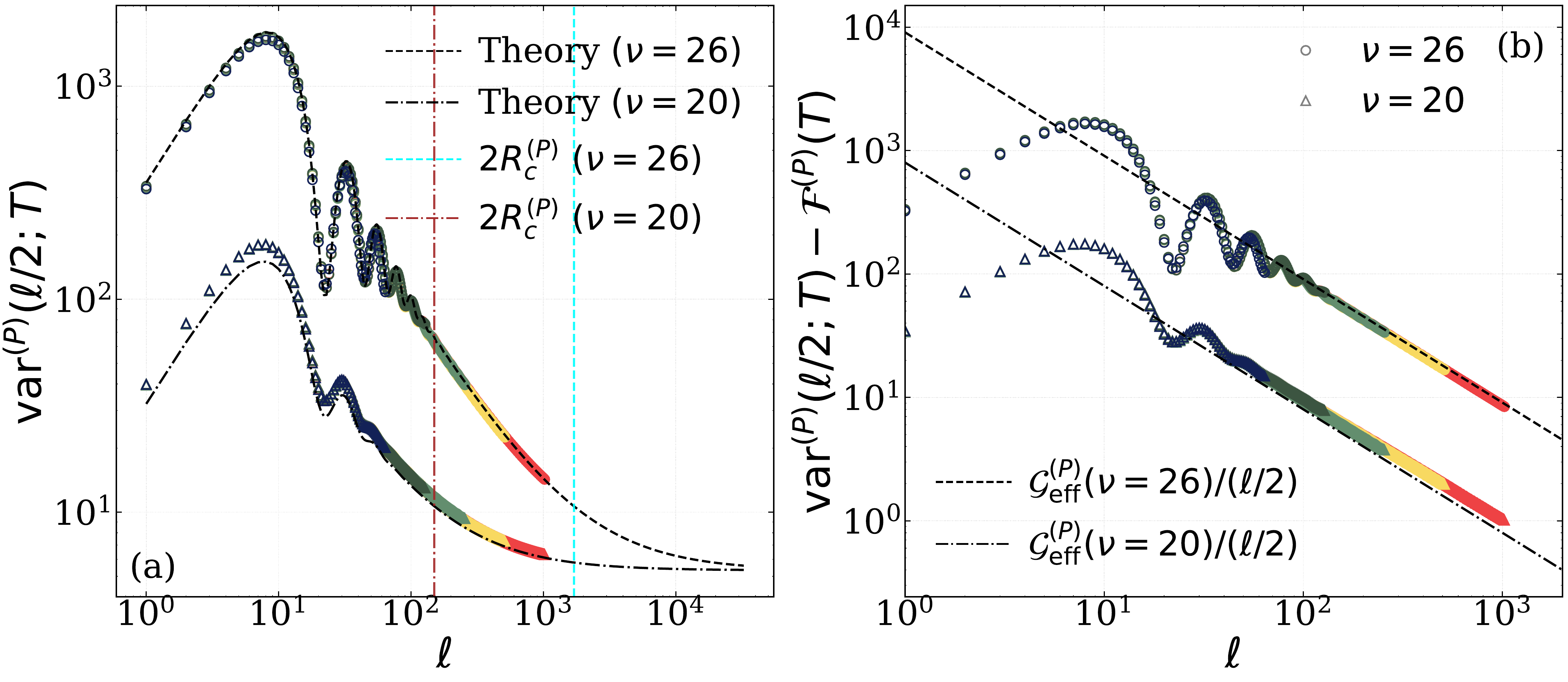}
    \caption{Subsystem variance of time-integrated prey density ($X=P$)
    at $T=4096$, for $\nu/\mu = 20$ ($R_c^{(P)} \approx 75$) and
    $\nu/\mu = 26$ ($R_c^{(P)} \approx 850$). Symbols: simulation
    data for $L = 128$--$2048$ (color codes as in
    Fig.~\ref{fig:prey_T32_T1}); dashed lines: LNA predictions with
    no fitting parameters. (a)~${\rm var}^{(P)}(\ell/2;\,T)$ converged
    to $\overline{{\rm var}}^{(P)}(\ell/2)$; vertical lines mark
    $\ell = 2R_c^{(P)}$. (b)~Subleading $1/\ell$ decay after
    subtracting $\mathcal{F}^{(P)}$, quantitatively captured by the
    LNA with no intrinsic cutoff.}
    \label{fig:variance_of_Prey_T4096}
\end{figure}

In Fig.~\ref{fig:variance_of_Prey_T4096}(a), we show
${\rm var}^{(P)}(\ell/2;\,T)$ at $T=4096$, for which the simulation
data have converged to the analytical LNA prediction at
$T \to \infty$, $\overline{{\rm var}}^{(P)}(\ell/2)$, for both
diffusivity ratios; the simulation time required for convergence
depends on model details and on proximity to the Turing instability.
The crossover scale $R_c^{(P)}$, defined by
$\overline{{\rm var}}^{(P)}(\ell/2) \simeq \mathcal{F}^{(P)}$ for
$\ell \gg 2R_c^{(P)}$, is $R_c^{(P)} \approx 75$ at $\nu/\mu = 20$
and $R_c^{(P)} \approx 850$ at $\nu/\mu = 26$ --- an
order-of-magnitude increase as the instability is approached. In
simulation, $\mathcal{F}^{(P)}$ is obtained from the structure
factor at $q=0$ of the time-integrated configurations, which
converge to its $T \to \infty$ value, as given in
Eq.~\eqref{eq:total_fluctuations}. Analytically, the $T \to \infty$
structure factor $\overline{S}_{q}^{(X)}$ is computed from the
dynamic power spectrum, as given in
Eq.~\eqref{eq:structure_factor_time_integrated_configs}. Substituting
$\overline{S}_{q}^{(X)}$ into Eq.~\eqref{eq:var_subsystem}, we
estimate $\mathcal{G}_{\rm eff}^{(X)}$ from the gradient of
$\overline{\rm var}^{(X)}(\ell/2) - \mathcal{F}^{(X)}$ for
$\ell \gg \ell^*$. Panel~(b) isolates the $1/\ell$ decay by
subtracting $\mathcal{F}^{(P)}$; the LNA quantitatively captures it
for $\ell > \ell^*$ across all system sizes. Crucially, this
$1/\ell$ decay has no intrinsic cutoff: only beyond
$\ell \sim 2R_c^{(P)}$ does $\mathcal{F}^{(P)}$ dominate and the
variance saturate, as confirmed by the systematic extension of the
decay range with $L$.

\begin{figure}
    \flushleft
    \includegraphics[width=\linewidth]{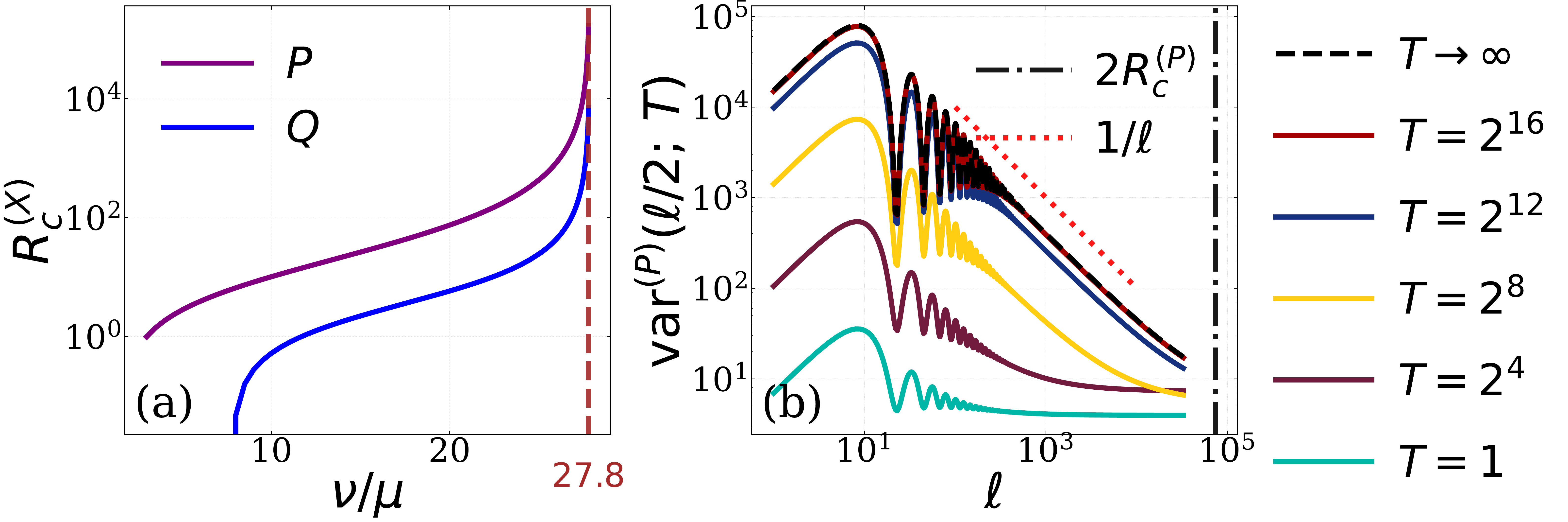}
    \caption{(a) Crossover radius $R_c^{(X)}$ versus $\nu/\mu$,
    computed from the LNA, diverges as the deterministic Turing
    instability ($\nu/\mu = 27.8$, dashed vertical line) is approached,
    with the prey (activator) diverging more strongly than the
    predator (inhibitor).
    (b) Analytically obtained, the intensive variance of prey at $\nu/\mu = 27.7$, as a
    function of subsystem size $\ell$, for integration times
    $T = 1, 16, 256, 4096, 65536$ and $T \to \infty$ (dashed black);
    the vertical dot-dashed line marks the cutoff length
    $2R_c^{(P)}$. The diverging $R_c^{(P)}$ sets an increasingly
    wide hyperuniform-scaling regime $\ell \ll 2R_c^{(P)}$ as $T \to \infty$ (see~\cite{SuppMat} for the predator).}
    \label{fig:lc_and_H}
\end{figure}
Fig.~\ref{fig:lc_and_H}(a) shows the divergence of the cutoff length
scale as $\nu/\mu$ approaches the deterministic instability value
for the given set of reaction parameters. In
Fig.~\ref{fig:lc_and_H}(b), using the analytical LNA very close to
the instability ($\nu/\mu = 27.7$), we show how
${\rm var}^{(P)}(\ell/2, T)$ evolves with increasing $T$, decaying
almost as $1/\ell$ over an increasingly broader region as
$T \to \infty$, with the hyperuniform-scaling regime extending to
$R_c^{(P)} \sim \mathcal{O}(10^5)$ in natural units.

\textit{Summary and conclusions.---}We have shown that, near the deterministic Turing instability,
time-integrated stochastic Turing patterns exhibit effective
hyperuniformity over a spatial-window larger than the Turing length scale
$R^*$: the population variance per unit volume decays as $1/R$
within this window, while instantaneous configurations remain
Poisson-like beyond $R^*$. Temporal integration can therefore reveal
spatial patterns otherwise hidden in instantaneous snapshots: in the
stochastic Levin-Segel model, the predator (inhibitor) at moderate
diffusivity ratios shows no discernible Turing length scale in any
snapshot --- it emerges only through temporal integration. This
effective hyperuniformity is an inevitable consequence of stochastic
Turing pattern formation, not a fine-tuned coincidence. The
large-subsystem fluctuation per unit volume is pinned to a reaction noise floor by a
sum rule on the correlation function --- a nonequilibrium
counterpart of the grand-canonical fluctuation-compressibility
relation --- while the growing peak at $R^*$ near the instability
drives the crossover length $R_c^{(X)}$ to increase monotonically,
widening the hyperuniform scaling regime as patterns sharpen.

Our work quantitatively demonstrates the emergence of a large-scale
hyperuniform-scaling regime in a stochastic multi-species
reaction-diffusion system driven by demographic noise alone, with no
conservation law required, in contrast to most earlier systems with
some conserved quantity where hyperuniform spatial organization has
been found~\cite{Hexner2017Jan, Mukherjee2024Aug}. The key insight
is that this organization is hidden in instantaneous configurations
and becomes visible only through temporal integration, which we
believe opens a route to discovering such hidden structure in
biological and ecological systems.

%%%%%%%%%%%%%%%%%%%%%%%%%%%%%%%%%%%%%%%%%%%%%%%%%%%%%%%%%%%%%%%%%%%%%%%%
%%%%%%%%%%%%%%%%%%%%%%%%%%%%%%%%%%%%%%%%%%%%%%%%%%%%%%%%%%%%%%%%%%%%%%%%
%%%%%%%%%%%%%%%%%%%%%%%%%%%%%%%%%%%%%%%%%%%%%%%%%%%%%%%%%%%%%%%%%%%%%%%%
%%%%%%%%%%%%%%%%%%%%%%%%%%%%%%%%%%%%%%%%%%%%%%%%%%%%%%%%%%%%%%%%%%%%%%%%

%%%%%%%%%%%%%%%%%%%%%%%%%%%%%%%%%%%%%%%%%%%%%%%%%% 
%%%%%%%%%%%%%%%%%%%%%%%%%%%%%%%%%%%%%%%%%%%%%%%%%% 
%%%%%%%%%%%%%%%%%%%%%%%%%%%%%%%%%%%%%%%%%%%%%%%%%% 

\textit{Acknowledgments.---} This work was supported by grants from the National Science and Technology Council, Taiwan (Grant No. NSTC 111-2112-M-001-027-MY3 and 114-2112-M-001-062) and Academia Sinica Career Development Award (Project No. AS-CDA-114-M02). AM acknowledges Punyabrata Pradhan and Murray Skolnick for useful discussions and Carlo Vanoni, Rituparna Mandal, and Sayani Chatterjee for reading the manuscript.

\bibliographystyle{apsrev4-2}
\bibliography{STP_Hyperuniformity_Letter_Refs}

\clearpage
\onecolumngrid

\setcounter{secnumdepth}{3}
\setcounter{equation}{0}
\setcounter{figure}{0}
\setcounter{section}{0}
\renewcommand{\theequation}{S\arabic{equation}}
\renewcommand{\thefigure}{S\arabic{figure}}
\renewcommand{\thesection}{S\Roman{section}}

\begin{center}
\textbf{\large Supplemental Material:\\[4pt]
Effective hyperuniformity in time-integrated stochastic Turing patterns}
\end{center}
\vspace{1em}

\section{Linear noise approximation (LNA) for a generic two-species stochastic reaction-diffusion model}

We consider a generic two-species stochastic reaction-diffusion process 
on a discrete $d$-dimensional hypercubic lattice with periodic boundary 
conditions, where each site $\vr$ has an internal volume $V$ and hosts 
non-negative integer populations $N^{(X)}_{\vr}(t)$, with $X=P$ (prey) 
and $X=Q$ (predator).
Each onsite reaction $\alpha$ generates a change 
$l_{\vr}^{(X, \alpha)}$ in the population of species $X$ with a 
transition propensity $W^{(\alpha)}_{\vr}$. Local populations also 
evolve due to random hopping of particles of species $X$ from site 
$\vr$ to $\vr'$ with propensity $\mathcal{W}^{(X)}_{\vr \vr'}$. 
In this letter, we only consider models with purely local reactions 
(excluding non-local processes that can trigger stochastic traveling 
waves~\cite{Biancalani2011Aug}), where both propensities satisfy the 
scaling relations $W^{(\alpha)}_{\vr}\br{N^{(X)}} = V 
w^{(\alpha)}_{\vr}\br{N^{(X)}/V}$ and 
$\mathcal{W}_{\vr\vr'}^{(X)}\br{N^{(X)}} = V 
\tilde{w}_{\vr\vr'}^{(X)}\br{N^{(X)}/V}$~\cite{Peralta2018Oct}.
Using the LNA~\cite{VanKampen2007, Lugo2008Nov}, the fluctuating local population $N_{\vr}^{(X)}$ can be written as
\begin{equation}
N_{\vr}^{(X)} \simeq V \rho_{\vr}^{(X)} + \sqrt{V} \xi_{\vr}^{(X)},
\end{equation}
the evolution equation 
of the local mean density $\rho_{\vr}^{(X)}$ is given by
\begin{equation}
  \label{eq:generic_deterministic_eqs}
  \begin{aligned}
    \frac{d \rho^{(X)}_{\vr}}{dt} = \sum_\alpha w^{(\alpha)}_{\vr}  
    l_{\vr}^{(X, \alpha)} +
    \sum_{\vr' \neq \vr} \br{ \tilde{w}^{(X)}_{\vr'\vr} - 
    \tilde{w}^{(X)}_{\vr\vr'} }, 
  \end{aligned}
\end{equation}
and the 
Langevin equation for the fluctuations $\xi_{\vr}^{(X)}$ by
\begin{equation}
  \label{eq:steady_state_langevin_eqn_form}
  \begin{aligned}
    \frac{d \xi_{\vr}^{(X)}}{dt} = \sum_{X'} \mathcal{R}^{(X X')} 
    \xi_{\vr}^{(X')} + \sum_{X'} \mathcal{D}^{(X X')} 
    \sum_{\vr'} \Delta_{\vr\vr'} \xi_{\vr'}^{(X')}  
    + \eta_{\vr}^{(X)}.
  \end{aligned}
\end{equation}
In the limit $V \rightarrow \infty$, the fluctuations in $N_{\vr}^{(X)}$ are completely determined by the dynamics of $\xi_{\vr}^{(X)}$.
The reaction coefficients $\mathcal{R}^{(X X')}$ and diffusion 
coefficients $\mathcal{D}^{(X X')}$ are given by
\begin{equation}
  \label{eq:generic_reactionand_diffusion_coefficients}
  \begin{aligned}
    \mathcal{R}^{(X X')} =& \sum_\alpha l_{\vr}^{(X, \alpha)} 
    \frac{\partial}{\partial \rho_{\vr}^{(X')}} 
    w^{(\alpha)}_{\vr}, \\
    \mathcal{D}^{(X X')} =& \frac{\partial}{\partial \rho_{\vr}^{(X')}}
    \tilde{w}_{\vr \vr'}^{(X)} - \frac{\partial}{\partial 
    \rho_{\vr}^{(X')}} \tilde{w}_{\vr'\vr}^{(X)};
  \end{aligned}
\end{equation}
the derivatives are evaluated at the homogeneous 
steady state, with the mean density $\rho^{(X)}$ of each species 
spatially uniform and satisfying
\begin{equation}
    \sum_\alpha w^{(\alpha)} l^{(X, \alpha)} = 0
\end{equation} 
for all $X$.

The noise strength in Fourier space, $B^{(X X')}$, 
is defined as
\begin{equation}
  \label{eq:noise_correlation_definition}
  \frac{\abr{\tilde{\eta}_{\vq}^{(X)}(\omega) 
\tilde{\eta}_{\vq'}^{(X')}(\omega')}}{2\pi L^d} = B^{(X X')} \delta_{\vq+\vq', 0}\, \delta(\omega + \omega'),
\end{equation}
where, by ignoring the weaker contribution from diffusion,
\begin{equation}
    \label{eq:noise_strength_formula}
    B^{(X X')} = \sum_\alpha 
  w^{(\alpha)} l^{(X, \alpha)} l^{(X', \alpha)};
\end{equation}
the corresponding Fourier transforms are defined as
\begin{equation}
  \label{eq:fourier_convention}
  \begin{aligned}
    \tilde{f}_\vq(\omega) &= \sum_{\vr} \int dt \, f_{\vr}(t) \, 
    e^{\iu \vq \cdot \vr} e^{\iu \omega t}, \\
    f_\vr(t) &= \frac{1}{2\pi L^d} \sum_{\vq} \int d\omega \, 
    \tilde{f}_\vq(\omega) \, e^{-\iu \vq \cdot \vr} 
    e^{-\iu \omega t},
  \end{aligned}
\end{equation}
with $\vr \in \{0,\ldots,L-1\}^d$ and $\vq = \frac{2\pi}{L}(n_1,\ldots,n_d)$, 
$n_a \in [0,L-1]$. We dropped the subleading diffusive 
contribution $2\tilde{w}^{(X)} \delta_{X,X'} \lambda_{\vq}$ from $B^{(XX')}$, 
where $\lambda_{\vq} = \sum_{a=1}^d 2(1-\cos{q_a}) 
\approx \abs{\vq}^2$ for $\abs{\vq}\rightarrow 0$ is the eigenvalue 
of the $d$-dimensional discrete Laplacian.

\subsection{Dynamic structure factor}
Eq.~\eqref{eq:steady_state_langevin_eqn_form} in Fourier space can be solved as
\begin{subequations}
\begin{align}
  \tilde{\xi}_{\vq}^{(X)}(\omega) = \sum_{X'} 
  \mathcal{M}^{(XX')}_{\vq}(\omega)\, 
  \tilde{\eta}_{\vq}^{(X')}(\omega),
  \label{eq:solution_of_fluctuations_in_fourier_space} 
\end{align}
and the dynamic structure factor $S_{\vq}^{(X)}(\omega)$, defined through
\begin{align}
     \frac{\abr{\tilde{\xi}_{\vq}^{(X)}(\omega)\, \tilde{\xi}_{\vq'}^{(X)}(\omega')}}{2\pi L^d} 
  = \delta_{\vq+\vq',0}\, \delta(\omega+\omega')\, 
    S_{\vq}^{(X)}(\omega),
  \label{eq:correlation_of_fluctuations_in_fourier_space}
\end{align}
then follows, using Eq.~\eqref{eq:solution_of_fluctuations_in_fourier_space}, as
\begin{equation}
  S_{\vq}^{(X)}(\omega) = \sum_{X',X''} 
  \mathcal{M}^{(XX')}_{\vq}(\omega)\, B^{(X'X'')}\, 
  \big[\mathcal{M}^{(XX'')}_{\vq}(\omega)\big]^{*}.
  \label{eq:dynamic_power_spectrum}
\end{equation}
\end{subequations}
The response matrix $\mathcal{M}_{\vq}(\omega)$ is
\begin{equation}
  \label{eq:rd_response_function}
  \mathcal{M}_{\vq}(\omega) = \left[ -\iu \omega\, \mathbb{I} 
  - \left(\mathcal{R} - \mathcal{D} \lambda_{\vq}\right)\right]^{-1}.
\end{equation}
We are particularly interested in the $\omega = 0$, $\abs{\vq} \to 0$ limit, where we can expand the response matrix as
\begin{equation}
  \label{eq:rd_response_function_small_q_zero_omega}
  \mathcal{M}_{\vq}(0) = -\br{\mathcal{R} - \mathcal{D} \lambda_{\vq}}^{-1} 
  \simeq - \br{\mathcal{R}^{-1} + \mathcal{R}^{-1}\mathcal{D} 
    \mathcal{R}^{-1} \lambda_{\vq} + \cdots},
\end{equation}
and write Eq.~\eqref{eq:dynamic_power_spectrum} as
\begin{equation}
  \label{sup:eq:final_zero_omega_small_q_spectrum}
    S^{(X)}_{\vq}(0) \simeq \mathcal{F}^{(X)} + 
    \mathcal{G}^{(X)} \abs{{\vq}}^2 + \cdots,
\end{equation}
where, using Eq.~\eqref{eq:noise_strength_formula},
\begin{subequations}
\begin{align}
\label{eq:leading_term}
\mathcal{F}^{(X)} &= \sum_{X',X''} (\mathcal{R}^{-1})^{(XX')}(\mathcal{R}^{-1})^{(XX'')} \sum_\alpha w^{(\alpha)} l^{(X', \alpha)} l^{(X'', \alpha)}, \\
\label{eq:subleading_term}
\mathcal{G}^{(X)} &= 2\sum_{X',X''} (\mathcal{R}^{-1})^{(XX')} (\mathcal{R}^{-1}\mathcal{D} \mathcal{R}^{-1})^{(XX'')} \sum_\alpha w^{(\alpha)} l^{(X', \alpha)} l^{(X'', \alpha)}.
\end{align}
\end{subequations}
$\mathcal{F}^{(X)}$ is the zero-frequency fluctuation of the total 
space-integrated population of species $X$, obtained by setting 
$|\vq| = 0$, independent of diffusion, and
\begin{equation}
    \label{sup:eq:pattern_formation_condition}
    \mathcal{G}^{(X)} > 0
\end{equation}
implies the stochastic Turing pattern-formation criterion of 
Ref.~\cite{Butler2011Jul}.

\section{Population fluctuation of time-integrated configurations in spherical windows}
The time-integrated population is defined as
\begin{equation}
    \Xi_\vr^{(X)}(T) = \int_0^T\, dt\, \xi_\vr^{(X)}(t).
    \label{eq:integrated_config}
\end{equation}
The time-integrated population within a spherical window
of radius $R$ centered at $\vr_0$ is
\begin{equation}
    \Xi^{(X)}(R;T) = \sum_\vr \Xi_\vr^{(X)}(T)\, w(\vr-\vr_0;R),
\end{equation}
where $w(\vr-\vr_0;R) = \mathbf{1}[|\vr-\vr_0|\leq R]$ is the spherical window indicator
on the hypercubic lattice and $v_1(R) = \sum_\vr w(\vr-\vr_0;R)$ is the window volume,
i.e.\ the number of lattice sites it encloses (in the continuum,
$v_1(R) = \pi^{d/2} R^d / \Gamma(d/2+1)$). We define the variance per unit volume and unit time as
\begin{equation}
    {\rm var}^{(X)}(R;T) = \frac{1}{v_1(R)\,T}\,\abr{\sbr{\Xi^{(X)}(R;T)}^2},
    \label{sup:eq:var_def}
\end{equation}
which is intensive and finite as $T\to\infty$.

\subsection{Structure factor of time-integrated configurations}
The Fourier transform of the time-integrated population fluctuation of species $X$ over $[0,T]$ is
\begin{equation}
    \tilde\Xi_\vq^{(X)}(T) = \int_0^T dt\,\tilde\xi_\vq^{(X)}(t)
    = \frac{1}{2\pi}\int d\omega\,\tilde\xi_\vq^{(X)}(\omega)\,W_T(\omega),
    \qquad
    W_T(\omega) = \int_0^T dt\,e^{-\iu\omega t}
    = \frac{1-e^{-\iu\omega T}}{\iu\omega},
    \label{sup:eq:windowed_transform}
\end{equation}
with window factor satisfying $W_T(-\omega) = W_T(\omega)^*$ and
\begin{equation}
    \abs{W_T(\omega)}^2 = W_T(\omega)\,W_T(-\omega) = \frac{2-2\cos\omega T}{\omega^2}.
    \label{sup:eq:window_kernel}
\end{equation}
Using the dynamic structure factor [Eq.~\eqref{eq:dynamic_power_spectrum}],
the structure factor of the time-integrated configurations $S_\vq^{(X)}(T)$ is
\begin{equation}
    \begin{aligned}
        S_\vq^{(X)}(T) &\equiv \frac{1}{L^d T}\,\abr{\abs{\tilde\Xi_\vq^{(X)}(T)}^2} \\
        &= \frac{1}{L^d T}\,\frac{1}{\br{2\pi}^2}\int d\omega\,d\omega'\,
        W_T(\omega)\,W_T(\omega')\,
        \abr{\tilde\xi_\vq^{(X)}(\omega)\,\tilde\xi_{-\vq}^{(X)}(\omega')} \\
        &= \frac{1}{2\pi T}\int d\omega\,d\omega'\,
        W_T(\omega)\,W_T(\omega')\,S_\vq^{(X)}(\omega)\,\delta(\omega+\omega') \\
        &= \frac{1}{2\pi T}\int d\omega\,S_\vq^{(X)}(\omega)\,\abs{W_T(\omega)}^2 \\
        &= \frac{1}{2\pi T}\int d\omega\, S_\vq^{(X)}(\omega)\,
        \frac{2-2\cos\omega T}{\omega^2}.
    \end{aligned}
    \label{sup:eq:Sq_T_def}
\end{equation}

\subsection{Variance in terms of $S_\vq^{(X)}(T)$}
Inserting the time-integrated definition of Eq.~\eqref{eq:integrated_config} into Eq.~\eqref{sup:eq:var_def} and Fourier transforming,
\begin{equation}
    \begin{aligned}
        {\rm var}^{(X)}(R;T) ={}& \frac{1}{v_1(R)\,T}
        \int_0^T dt \int_0^T dt'\,
        \sum_{\vr,\vr'} \abr{\xi_\vr^{(X)}(t)\,\xi_{\vr'}^{(X)}(t')}\,
        w(\vr-\vr_0;R)\,w(\vr'-\vr_0;R) \\
        ={}& \frac{1}{v_1(R)\,T}\,\frac{1}{\br{2\pi L^d}^2}
        \int_0^T dt \int_0^T dt'
        \sum_{\vr,\vr'}\sum_{\vq,\vq'}\int d\omega\,d\omega'\,
        \abr{\tilde\xi_\vq^{(X)}(\omega)\,\tilde\xi_{\vq'}^{(X)}(\omega')} \\
        &\times e^{-\iu\vq\cdot\vr}\,e^{-\iu\vq'\cdot\vr'}\,
        e^{-\iu\omega t}\,e^{-\iu\omega' t'}\,
        w(\vr-\vr_0;R)\,w(\vr'-\vr_0;R).
    \end{aligned}
\end{equation}
Using Eq.~\eqref{eq:correlation_of_fluctuations_in_fourier_space}, we can further write,
\begin{equation}
    {\rm var}^{(X)}(R;T) = \frac{1}{v_1(R)\,T}\,\frac{1}{2\pi L^d}
    \sum_{\vr,\vr'} w(\vr-\vr_0;R)\,w(\vr'-\vr_0;R)
    \sum_{\vq} e^{-\iu\vq\cdot(\vr-\vr')}
    \int d\omega\, S_\vq^{(X)}(\omega)\,\frac{2-2\cos\omega T}{\omega^2}.
\end{equation}
By Eq.~\eqref{sup:eq:Sq_T_def} the frequency integral is exactly $2\pi T\,S_\vq^{(X)}(T)$, which cancels the $1/(2\pi T)$ prefactor,
\begin{equation}
    {\rm var}^{(X)}(R;T) = \frac{1}{v_1(R)\,L^d}
    \sum_{\vq} S_\vq^{(X)}(T)
    \sum_{\vr,\vr'} w(\vr-\vr_0;R)\,w(\vr'-\vr_0;R)\,
    e^{-\iu\vq\cdot(\vr-\vr')}.
\end{equation}
Introducing the window form factor,
\begin{equation}
    \tilde w_\vq(R) = \sum_\vr w(\vr;R)\,e^{-\iu\vq\cdot\vr},
    \qquad \abs{\tilde w_\vq(R)}^2 = 
    \sum_{\vr,\vr' \in \Omega(R)}\,
    e^{-\iu\vq\cdot(\vr-\vr')},
\end{equation}
where $\Omega(R) = \{\vr : |\vr|\leq R\}$ is the set of lattice sites
enclosed by the window and $v_1(R) = |\Omega(R)|$ is their number,
approaching the continuum ball volume
$v_1(R) \to \pi^{d/2}R^d/\Gamma(d/2+1)$ only for $R\gg 1$.
Assuming translation invariance, so the two-point statistics are
independent of the origin $\vr_0$, the variance takes the local-variance
form for the time-integrated configurations,
\begin{equation}
    {\rm var}^{(X)}(R;T)
    = \frac{1}{L^d}\sum_\vq
    S_\vq^{(X)}(T)\, \tilde{\alpha}_\vq(R);\qquad
    \tilde{\alpha}_\vq(R) = \frac{\abs{\tilde w_\vq(R)}^2}{v_1(R)},
    \label{sup:eq:var_Sq_T}
\end{equation}
where $\tilde\alpha_\vq(R)$ is the scaled window intensity, the squared
window form factor normalized by its volume, which couples the structure
factor to the window geometry; this is the time-integrated analogue of the
variance formula for a point process~\cite{Torquato2003Oct}.

\subsection{Asymptotic formula for $R \gg 1$}
In the large-$R$ limit, we approximate the lattice sum in the window form factor by a continuum integral,
\begin{equation}
    \tilde{w}_\vq(R) \approx R^d \int_{\vr' \in \Omega(1)} d\vr'\, e^{-\iu R \vq \cdot \vr'}
    = R^d \int_0^1 \abs{\vr'}^{d-1}\, d\abs{\vr'}\, \frac{2\pi^{(d-1)/2}}{\Gamma\!\left(\frac{d-1}{2}\right)}
    \int_0^\pi \sin^{d-2}\!\theta\; e^{-\iu R\abs{\vq}\abs{\vr'}\cos\theta}\, d\theta,
    \label{sup:eq:approx_window_form_factor}
\end{equation}
where $\theta = \angle(\hat{\vq},\hat{\vr}')$ and the prefactor is the surface area of the unit $(d-2)$-sphere. Using the standard identity~\cite{NIST}
\begin{equation}
    \frac{2\pi^{(d-1)/2}}{\Gamma\!\left(\frac{d-1}{2}\right)}
    \int_0^\pi \sin^{d-2}\!\theta\; e^{-\iu z\cos\theta}\, d\theta
    = \frac{(2\pi)^{d/2}}{z^{(d-2)/2}}\,J_{(d-2)/2}(z),
\end{equation}
with $z = R\abs{\vq}\abs{\vr'}$, and the Bessel integral identity $\int_0^1 r^{\nu+1}J_\nu(ar)\,dr = J_{\nu+1}(a)/a$ with $\nu=(d-2)/2$ and $a=R\abs{\vq}$, the radial integral evaluates to give
\begin{equation}
    \tilde{w}_\vq(R) \approx \frac{(2\pi R)^{d/2}}{\abs{\vq}^{d/2}}\,J_{d/2}(R\abs{\vq}), \qquad R \gg 1,
\end{equation}
and hence
\begin{equation}
    \tilde{\alpha}_\vq(R) = \frac{\abs{\tilde{w}_\vq(R)}^2}{v_1(R)}
    \approx \frac{2^d\pi^{d/2}\,\Gamma(d/2+1)}{\abs{\vq}^d}\,J_{d/2}^2(R\abs{\vq}),
    \qquad R \gg 1,
    \label{sup:eq:alpha_bessel}
\end{equation}
where $v_1(R) = \pi^{d/2}R^d/\Gamma(d/2+1)$ is the volume of the $d$-ball~\cite{Torquato2003Oct}, recovered in this limit. For $d=2$ this reduces to $\tilde{\alpha}_\vq(R) \approx 4\pi J_1^2(R\abs{\vq})/\abs{\vq}^2$.

\subsection{Variance formula in the limit $T \to \infty$}
In the limit $T\to\infty$, the kernel $\tfrac{1}{2\pi T}|W_T(\omega)|^2 \to \delta(\omega)$,
and hence from Eq.~\eqref{sup:eq:Sq_T_def},
\begin{equation}
    \overline{S}_\vq^{(X)} \equiv \lim_{T\to\infty} S_\vq^{(X)}(T) = S_\vq^{(X)}(0),
\end{equation}
so the time-integrated structure factor converges to the zero-frequency slice of the dynamic power spectrum, from which
\begin{equation}
    \overline{\rm var}^{(X)}(R) \equiv \lim_{T \to \infty} {\rm var}^{(X)}(R;T)
    = \frac{1}{L^d} \sum_\vq \overline{S}_\vq^{(X)}\, \tilde{\alpha}_\vq(R).
\end{equation}

\section{Stochastic Levin-Segel model}

We consider a reactive and diffusive population of prey and predator 
species in a periodic one-dimensional lattice of size $L$. Each lattice 
point $i = 0, 1, 2, \ldots, L-1$ has a well-mixed internal volume, 
parameterized by $V$, which can be populated by both prey and predators, 
$N^{(P)}_i \in \{0, 1, 2, \ldots\}$ and $N^{(Q)}_i \in \{0, 1, 2, 
\ldots\}$. The 
reaction and diffusion dynamics are governed by 
Eq.~\eqref{eq:LS-stochastic-model}, where $j$ denotes a neighboring 
lattice site, $z$ is the coordination number, and $\mu$, $\nu$ are the 
diffusion rates for prey and predator, respectively.
The propensities of each reaction and the corresponding changes in each 
species are given in Table~\ref{tab:reactions}.
\begin{figure}[htbp]
\centering
\begin{minipage}{0.46\linewidth}
  \centering
  \begin{equation}
    \label{eq:LS-stochastic-model}
    \begin{aligned}
        P_i &\xrightarrow{b} 2P_i, \\
        2P_i &\xrightarrow{e/V} 3P_i, \\
        P_i + Q_i &\xrightarrow{p/V} 2Q_i, \\
        2Q_i &\xrightarrow{d/V} Q_i, \\
        P_i &\xrightarrow{z\mu} P_j, \quad
        Q_i \xrightarrow{z\nu} Q_j.
    \end{aligned}
  \end{equation}
\end{minipage}
\hfill
\begin{minipage}{0.50\linewidth}
  \centering
  \captionof{table}{Scaled rates and stoichiometric changes for the 
  four onsite reactions.}
  \label{tab:reactions}
  \begin{tabular}{lccc}
  \hline
  Reaction & $w^{(\alpha)}$ & $l^{(P,\alpha)}$ & $l^{(Q,\alpha)}$ \\
  \hline
  $P \to 2P$ & $b\bar{\rho}^{(P)}$ & $+1$ & $0$ \\
  $2P \to 3P$ & $e(\bar{\rho}^{(P)})^2$ & $+1$ & $0$ \\
  $P + Q \to 2Q$ & $p\bar{\rho}^{(P)}\bar{\rho}^{(Q)}$ & $-1$ & $+1$ \\
  $2Q \to Q$ & $d(\bar{\rho}^{(Q)})^2$ & $0$ & $-1$ \\
  \hline
  \end{tabular}
\end{minipage}
\end{figure}
$\bar{\rho}^{(P)}$ and $\bar{\rho}^{(Q)}$ are the steady-state solution 
of the following deterministic equations satisfied by the local prey and 
predator densities, $\rho^{(X)}_i = \abr{N_i^{(X)}} / V$,
\begin{equation}
    \label{eq:levin-segel_deterministic_eqn}
   \begin{aligned}
        \frac{\partial \rho_i^{(P)}}{\partial t} &= b \rho_i^{(P)} 
        + e {\rho_i^{(P)}}^2 - p \rho_i^{(P)} \rho_i^{(Q)} 
        + \mu \sum_j \nabla^2_{ij} \rho_j^{(P)}, \\
        \frac{\partial \rho_i^{(Q)}}{\partial t} &= p \rho_i^{(Q)} \rho_i^{(P)} 
        - d {\rho_i^{(Q)}}^2 + \nu \sum_j \nabla^2_{ij} \rho_j^{(Q)},
   \end{aligned}
\end{equation}
with homogeneous steady-state densities
\begin{align}
\label{eq:LS_without_capacity_sol}
   \bar{\rho}^{(P)} = \frac{bd}{p^2-de}, \quad 
   \bar{\rho}^{(Q)} = \frac{bp}{p^2-de}.
\end{align}
$\nabla^2$ is the discrete Laplacian, defined as $\sum_j \nabla^2_{ij} \rho_j^{(X)} = \br{\rho_{i-1}^{(X)} + \rho_{i+1}^{(X)} - 2\rho_i^{(X)}}$.

\subsection{Reaction and diffusion matrices}
Using Eq.~\eqref{eq:generic_reactionand_diffusion_coefficients}, we write the reaction and diffusion matrices as
\begin{align}
  \mathcal{R} = \frac{b}{p^2-de} 
  \begin{pmatrix}
    de & -dp \\
    p^2 & -dp
  \end{pmatrix}, \quad
  \mathcal{D} =
  \begin{pmatrix}
    \mu & 0 \\
    0 & \nu
  \end{pmatrix},
\end{align}
with $\det(\mathcal{R}) = b^2dp/(p^2-de)$, yielding the inverse
\begin{align}
  \label{eq:ls_model_inverse_matrices}
  \mathcal{R}^{-1} = \frac{1}{b}
  \begin{pmatrix}
    -1 & 1 \\
    -p/d & e/p
  \end{pmatrix}.
\end{align}
We also compute $\mathcal{R}^{-1}\mathcal{D}\mathcal{R}^{-1}$, as it is needed for $\mathcal{G}^{(X)}$:
\begin{align}
\label{eq:RiDRi}
  \mathcal{R}^{-1}\mathcal{D}\mathcal{R}^{-1} = \frac{1}{b^2}
  \begin{pmatrix}
    \mu - \nu p/d & -\mu + \nu e/p \\
    \mu p/d - \nu e/d & -\mu p/d + \nu e^2/p^2
  \end{pmatrix}.
\end{align}

\subsection{Reaction noise matrix}
The components of the noise strengths, given by 
Eq.~\eqref{eq:noise_strength_formula}, can be written using 
Table~\ref{tab:reactions} as
\begin{equation}
  \label{eq:noise_components_LS}
  \begin{aligned}
    B^{(PP)} &= b\bar\rho^{(P)} + e(\bar\rho^{(P)})^2 
      + p\bar\rho^{(P)}\bar\rho^{(Q)}, \\
    B^{(PQ)} &= B^{(QP)} = -\,p\bar\rho^{(P)}\bar\rho^{(Q)}, \\
    B^{(QQ)} &= p\bar\rho^{(P)}\bar\rho^{(Q)} + d(\bar\rho^{(Q)})^2,
  \end{aligned}
\end{equation}
where only the third reaction $P+Q\to 2Q$ contributes to the 
off-diagonal $B^{(PQ)}$, since it is the only reaction that changes 
both species. Substituting the steady-state densities from 
Eq.~\eqref{eq:LS_without_capacity_sol}, we write the complete noise-strength matrix as
\begin{equation}
  \label{eq:noise_matrix_LS}
  B = \frac{b^2 d p^2}{(p^2-de)^2}
  \begin{pmatrix}
    2 & -1 \\
    -1 & 2
  \end{pmatrix}.
\end{equation}

\subsection{Computation of $\mathcal{F}^{(X)}$}

Substituting Eqs.~\eqref{eq:ls_model_inverse_matrices} and 
\eqref{eq:noise_matrix_LS} into Eq.~\eqref{eq:leading_term}, we obtain the following expressions of $\mathcal{F}^{(X)}$.

\paragraph{Prey ($X=P$):}
\begin{align}
  \mathcal{F}^{(P)} 
  &= \sum_{X',X''} (\mathcal{R}^{-1})^{(PX')}
     (\mathcal{R}^{-1})^{(PX'')}\,B^{(X'X'')} \notag\\
  &= \frac{1}{b^2}\frac{b^2dp^2}{(p^2-de)^2}
    \bigl[(-1)^2\cdot 2 + 2(-1)(+1)(-1) + (+1)^2\cdot 2\bigr] 
    \notag\\
  &= \frac{6dp^2}{(p^2-de)^2}.
\end{align}

\paragraph{Predator ($X=Q$):}
\begin{align}
  \mathcal{F}^{(Q)} 
  &= \frac{1}{b^2}\frac{b^2dp^2}{(p^2-de)^2}
    \left[\left(-\frac{p}{d}\right)^2\cdot 2 
    + 2\!\left(-\frac{p}{d}\right)\!\left(\frac{e}{p}\right)(-1) 
    + \left(\frac{e}{p}\right)^2\!\cdot 2\right] \notag\\
  &= \frac{2dp^2}{(p^2-de)^2}
    \left[\frac{p^2}{d^2} + \frac{e}{d} + \frac{e^2}{p^2}\right]
    \notag\\
  &= \frac{2(d^2e^2+dep^2+p^4)}{d(p^2-de)^2}.
\end{align}

\subsection{Computation of $\mathcal{G}^{(X)}$}
Substituting Eqs.~\eqref{eq:ls_model_inverse_matrices},~\eqref{eq:RiDRi} and~\eqref{eq:noise_matrix_LS} into Eq.~\eqref{eq:subleading_term}, we obtain the following expressions of $\mathcal{G}^{(X)}$.

\paragraph{Prey ($X=P$):}
\begin{align}
  \mathcal{G}^{(P)}
  &= 2\sum_{X',X''}(\mathcal{R}^{-1})^{(PX')}
     (\mathcal{R}^{-1}\mathcal{D}\mathcal{R}^{-1})^{(PX'')}
     B^{(X'X'')} \notag\\
  &= \frac{2}{b}\cdot\frac{b^2dp^2}{(p^2-de)^2}\cdot\frac{1}{b^2}
  \Bigl[(-1)\bigl((\mu-\nu p/d)\cdot 2 
         + (-\mu+\nu e/p)\cdot(-1)\bigr) \notag\\
  &\qquad + (+1)\bigl((\mu-\nu p/d)\cdot(-1) 
         + (-\mu+\nu e/p)\cdot 2\bigr)\Bigr] \notag\\
  &= \frac{6p\,[\nu(p^2+de) - 2\mu dp]}{b(p^2-de)^2}.
\end{align}

\paragraph{Predator ($X=Q$):}
\begin{align}
  \mathcal{G}^{(Q)}
  &= \frac{2}{b}\cdot\frac{b^2dp^2}{(p^2-de)^2}\cdot\frac{1}{b^2}
  \Bigl[\!\left(-\tfrac{p}{d}\right)\!\bigl((\mu p/d-\nu e/d)\cdot 2 
        + (-\mu p/d+\nu e^2/p^2)\cdot(-1)\bigr) \notag\\
  &\qquad + \tfrac{e}{p}\bigl((\mu p/d-\nu e/d)\cdot(-1) 
        + (-\mu p/d+\nu e^2/p^2)\cdot 2\bigr)\Bigr] \notag\\
  &= \frac{2\,[\,2\nu e(p^4+dep^2+d^2e^2)
        - 3\mu p^3(p^2+de)\,]}{bdp(p^2-de)^2}.
\end{align}

\subsection{Final spectra at zero frequency and small wave vector limit}
Assembling $\mathcal{F}^{(X)}$ and $\mathcal{G}^{(X)}$ in Eq.~\eqref{sup:eq:final_zero_omega_small_q_spectrum},
$S_q^{(X)}(0) \simeq \mathcal{F}^{(X)} + \mathcal{G}^{(X)}q^2$ gives
\begin{equation}
  \begin{aligned}
    \label{eq:analytical_expression_of_spectrum_ls}
    S_{q}^{(P)}(0) &\simeq \frac{6dp^2}{(p^2-de)^2}
      + \frac{6p\,[\nu(p^2+de)-2\mu dp]}{b(p^2-de)^2}\,q^2, \\[6pt]
    S_{q}^{(Q)}(0) &\simeq \frac{2(d^2e^2+dep^2+p^4)}{d(p^2-de)^2}
      + \frac{2\,[\,2\nu e(p^4+dep^2+d^2e^2)
            -3\mu p^3(p^2+de)\,]}{bdp(p^2-de)^2}\,q^2.
  \end{aligned}
\end{equation}

\subsection{Results for predator population}
\begin{figure}
    \centering
    \includegraphics[width=\linewidth]{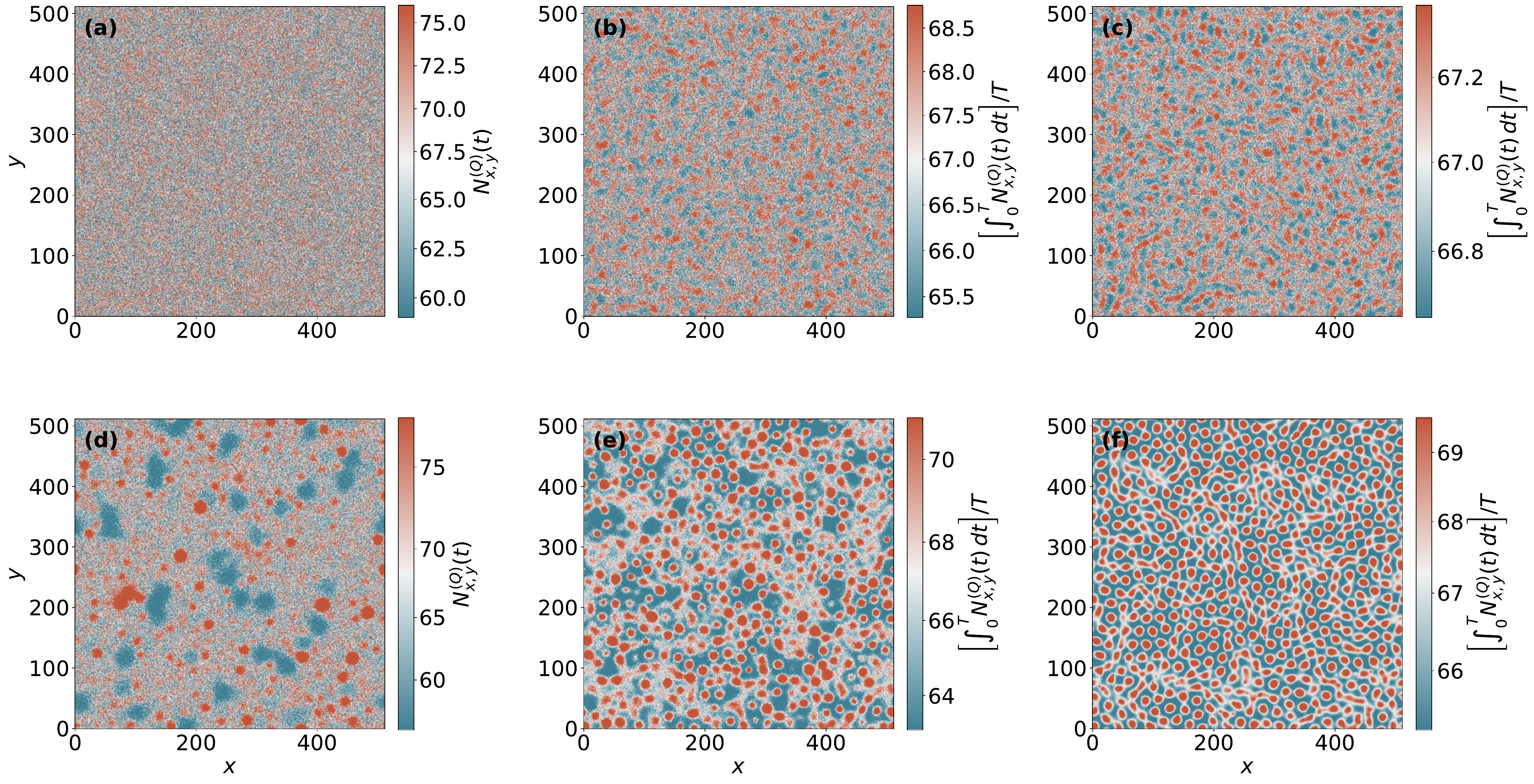}
    \caption{Spatial patterns of predators in the 2D Levin--Segel model on a
    $512 \times 512$ grid ($b = e = d = 0.5$, $p = 1$, $\mu = 1$, $V = 100$).
    (a--c) $\nu/\mu = 20$; (d--f) $\nu/\mu = 26$. Left panels (a,d):
    instantaneous steady-state configurations; middle and right panels
    (b,e) and (c,f): time-integrated over $T = 32$ and $T = 1024$
    respectively. The organized predator structure emerges with increasing
    integration time, absent in any single snapshot. The 2D simulation is
    shown for visualization only; all quantitative analyses use the 1D model.}
    \label{fig:predator_spatial}
\end{figure}

The instantaneous and time-integrated predator configurations in two
dimensions, analogous to Fig.~(1) for prey in the main text, are shown in
Fig.~\ref{fig:predator_spatial}; the upper panel is for $\nu/\mu = 20$ and
the lower for $\nu/\mu = 26$. At $\nu/\mu = 20$, the instantaneous
configuration [Fig.~\ref{fig:predator_spatial}(a)] looks almost random and
lacks any characteristic length scale [see $S_{\vq}^{(Q)}$,
Fig.~\ref{fig:structure_factors}(b)], and visible patterns appear only in
the time-integrated configurations [Fig.~\ref{fig:predator_spatial}(b),(c)].
Sharper predator patterns develop near the Turing instability at
$\nu/\mu = 26$ (lower panel), where large patches already form in the
instantaneous configuration [Fig.~\ref{fig:predator_spatial}(d)], and the
time-integrated configurations [Fig.~\ref{fig:predator_spatial}(e),(f)]
develop much sharper patterns than their away-from-instability counterparts.

\begin{figure}
    \includegraphics[width=\linewidth]{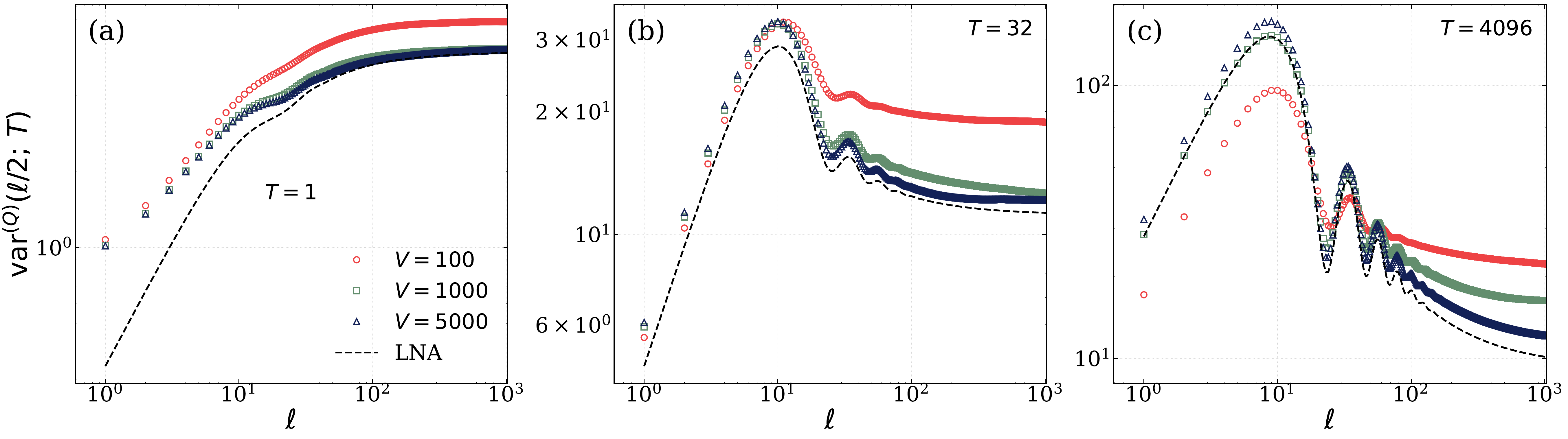}
    \caption{Intensive variance of time-integrated predator configurations,
    $\mathrm{var}^{(Q)}(\ell/2;T)$, versus subsystem size $\ell$ for
    $L=2048$, $\nu/\mu=26$ (near the instability), and $V=100$ (circles),
    $V=1000$ (squares), $V=5000$ (triangles). Panels (a),(b),(c) correspond
    to $T=1$, $32$, and $4096$. Open markers: stochastic simulations via the
    Next Reaction Method; dashed lines: Linear Noise Approximation (LNA).
    Agreement improves with increasing $V$, consistent with the LNA.}
    \label{fig:predator_var_nu26}
\end{figure}

We characterize the one-dimensional predator patterns at $\nu/\mu = 26$ by
the intensive variance $\mathrm{var}^{(Q)}(\ell/2;T)$ on a ring. In
Fig.~\ref{fig:predator_var_nu26}(a), for $T=1$ the variance shows no $1/\ell$
decaying regime; this appears only in the intermediate time-integrated
configuration $T=32$ [Fig.~\ref{fig:predator_var_nu26}(b)]. At $T=4096$
[Fig.~\ref{fig:predator_var_nu26}(c)], the peak at the Turing length scale
and the large-scale saturation value are well separated, signifying a
sufficiently developed pattern. For increasing $V$, the large-scale
fluctuations converge to their analytical value.

\begin{figure}
    \includegraphics[width=\linewidth]{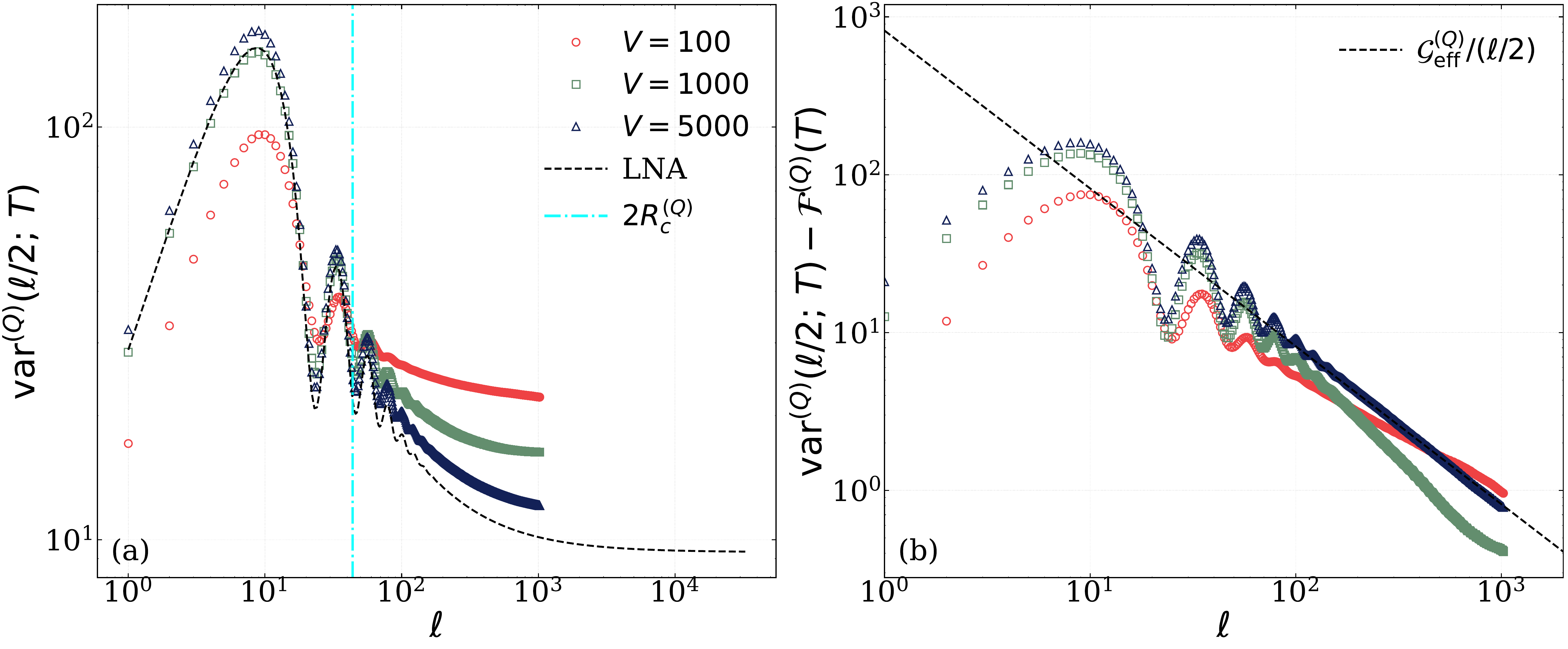}
    \caption{Intensive variance of time-integrated predator configurations
    for $L=2048$, $\nu/\mu=26$ (near the instability), at $T=4096$.
    (a) Raw variance $\mathrm{var}^{(Q)}(\ell/2;T)$ versus subsystem size
    $\ell$, for $V=100$ (circles), $V=1000$ (squares), $V=5000$ (triangles),
    compared against the LNA (dashed black line). The cyan dot-dashed vertical
    line marks $2R_c^{(Q)}$, beyond which the variance approaches its
    large-scale saturation value. (b) Variance with the saturation plateau
    subtracted, $\mathrm{var}^{(Q)}(\ell/2;T) - \mathcal{F}^{(Q)}(T)$,
    isolating the correlated contribution. The dashed line shows the
    predicted $\mathcal{G}_{\rm eff}^{(Q)}/(\ell/2)$ scaling, in good
    agreement for $\ell \gg 1$ and $V=5000$.}
    \label{fig:predator_varF_nu26}
\end{figure}

To isolate the correlated component, we subtract
$\mathcal{F}^{(Q)}(T) \equiv S^{(Q)}_{\vq=0}(T)$ at $T=4096$ from the
simulated variance [Fig.~\ref{fig:predator_varF_nu26}(a)]. The subtracted
variance [Fig.~\ref{fig:predator_varF_nu26}(b)] follows a $1/\ell$ decay.
Both panels are compared against the $T\to\infty$ LNA prediction, with
$V=5000$ showing very good agreement.

\subsection{Population fluctuations near the Turing instability}
\begin{figure}
    \includegraphics[width=\linewidth]{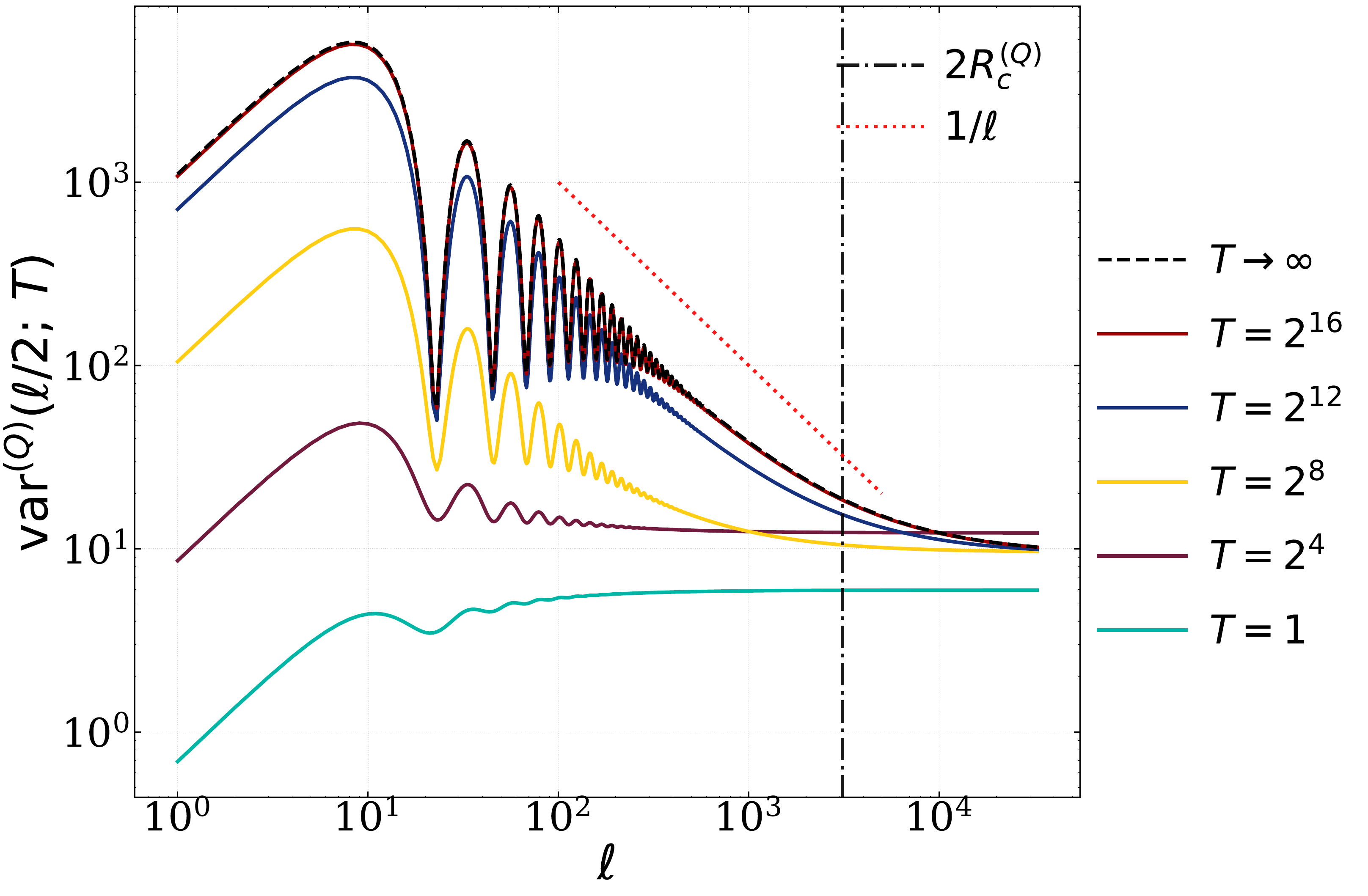}
    \caption{Intensive variance of time-integrated predator configurations,
    $\mathrm{var}^{(Q)}(\ell/2;T)$, in the 1D Levin--Segel model at
    $\nu/\mu = 27.7$ (near the instability), versus subsystem size $\ell$.
    Curves show the Linear Noise Approximation (LNA) predictions for
    integration times $T = 1, 16, 256, 4096, 65536$ and $T \to \infty$
    (dashed black). The vertical dot-dashed line marks the cutoff length
    $2R_c^{(Q)}$, beyond which the variance saturates to its $\ell \gg 1$
    limit.}
    \label{fig:predator_subsystem_fluct_nu27p7}
\end{figure}
We examine analytically how the predator subsystem variance evolves with
integration time $T$ even closer to the instability, at $\nu/\mu = 27.7$,
using the LNA prediction
(Fig.~\ref{fig:predator_subsystem_fluct_nu27p7}). For
$\ell \ll 2R_c^{(Q)}$, the variance increases monotonically with $T$,
approaching the $T \to \infty$ limit only once $T$ is large enough. Beyond $2R_c^{(Q)}$ the
variance saturates nonmonotonically with $T$. Within this intermediate region, predator population is showing an effectively-hyperuniform $1/\ell$ decay for a much broader range compared to the previous $\nu / \mu$ ratios.

\subsection{Comparison of structure factors near and away from the instability}
\begin{figure}
    \includegraphics[width=\linewidth]{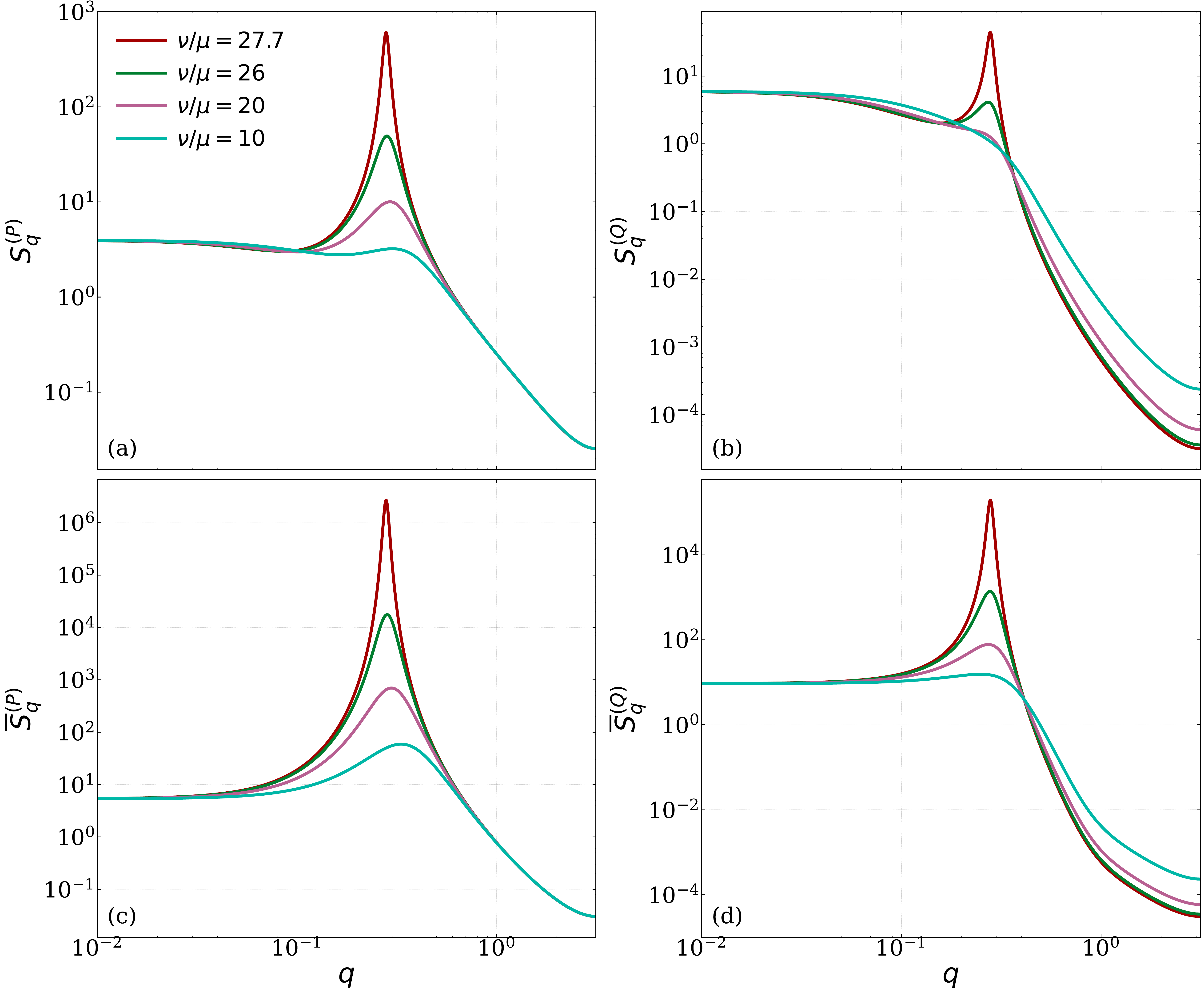}
    \caption{Structure factors of prey (left) and predator (right)
    populations in the 1D Levin--Segel model for
    $\nu/\mu = 27.7,\,26,\,20,\,10$, spanning the near-instability
    ($\nu/\mu = 26,\,27.7$) and away-from-instability ($\nu/\mu = 10,\,20$)
    regimes. Top row (a,b): instantaneous structure factor $S_q^{(X)}$;
    bottom row (c,d): time-integrated structure factor $\overline{S}_q^{(X)}$
    in the $T\to\infty$ limit. As $\nu/\mu$ increases toward the instability,
    the peak at the finite Turing wavenumber $q^*$ sharpens in both species,
    while $S_q^{(X)}\!\to\!\mathcal{F}^{(X)}$ remains finite as $q\to0$.
    Time integration suppresses the noise of single snapshots and
    isolates the slow $\omega=0$ modes, so the Turing peak stands out far
    more clearly in $\overline{S}_q^{(X)}$.}
    \label{fig:structure_factors}
\end{figure}
Figure~\ref{fig:structure_factors} compares the instantaneous and
time-integrated structure factors across the four values of $\nu/\mu$.
Near the instability ($\nu/\mu = 26,\,27.7$), a peak at the Turing
wavenumber $q^*$ is present already in the instantaneous structure factor
[panels (a),(b)] and sharpens further under time integration
[panels (c),(d)]. Away from the instability ($\nu/\mu = 10,\,20$) the
behavior is markedly different, and most pronounced for the predator: its
instantaneous structure factor $S_q^{(Q)}$ is featureless, with no peak
whatsoever [panel (b)], so the latent pattern is entirely masked by
noise in any single snapshot. The Turing peak emerges only in the
time-integrated structure factor $\overline{S}_q^{(Q)}$ in the $T\to\infty$
limit [panel (d)], where time integration acts as a low-pass filter on the
$\omega=0$ modes. The spatial organization thus persists even far from the
instability but, for the predator especially, is resolvable only after
temporal accumulation.
% ==============================

\end{document}